# Quantization of the Electromagnetic Field in Non-dispersive Polarizable Moving Media above the Cherenkov Threshold

*Mário G. Silveirinha*[*]

[(1)]*University of Coimbra, Department of Electrical Engineering – Instituto de Telecomunicações, Portugal, mario.silveirinha@co.it.pt*

**Abstract**

We quantize the macroscopic electromagnetic field in a system of non-dispersive polarizable bodies moving at constant velocities possibly exceeding the Cherenkov threshold. It is shown that in general the quantized system is unstable and neither has a ground state nor supports stationary states. The quantized Hamiltonian is written in terms of quantum harmonic oscillators associated with both positive and negative frequencies, such that the oscillators associated with symmetric frequencies are coupled by an interaction term that does not preserve the quantum occupation numbers. Moreover, in the linear regime the amplitudes of the fields may grow without limit provided the velocity of the moving bodies is enforced to be constant. This requires the application of an external mechanical force that effectively pumps the system.

---

[*] To whom correspondence should be addressed: E-mail: *mario.silveirinha@co.it.pt*

# I. Introduction

The quantization of the electromagnetic field in material media has been a topic of intense research in the framework of macroscopic quantum electrodynamics. In particular, quantization procedures have been developed for the cases of both nondispersive and frequency dispersive dielectrics, inhomogeneous structures, and even materials with time-dependent responses (see e.g. Refs. [1-5]). Quantized field theories are invariably constructed under the assumption that the system under analysis is *stable*, such that in the absence of loss the classical natural modes of oscillation of the electromagnetic field have infinite decay times and are characterized by a certain frequency of oscillation. Thus, the natural modes have a time variation of the form $e^{-i\omega t}$ with the frequency $\omega$ real valued. Each classical natural mode of oscillation is usually associated with a quantum harmonic oscillator. In case of dissipation, the lifetime $\tau = 1/(-2\lambda)$ of a natural mode is finite, and as a consequence the frequency of oscillation becomes complex valued, $\omega = \omega' + i\lambda$, with $\lambda < 0$, such that the time variation is of the form $e^{-i\omega' t} e^{\lambda t}$. Dissipative systems are not electromagnetically closed, and their quantization typically requires introducing a reservoir that interacts with the system.

There are obvious and logical difficulties in generalizing these ideas to the case of electromagnetically unstable systems. In such systems the electromagnetic fields can *grow* with time because of some system instability, so that the natural modes are described a oscillation frequency $\omega = \omega' + i\lambda$ with $\lambda > 0$. This scenario is precisely the playground for this work.

The occurrence of system instabilities may be at first hard to imagine without considering explicitly an electromagnetic "pump" or "source of radiation" in order that the natural modes can grow with time. We show that somewhat surprisingly such instabilities may occur in "wave closed systems" (in the sense that the wave energy and wave momentum are conserved) formed by moving *uncharged* bodies (modeled as continuous media and such that the planar surfaces are perfectly smooth) in case their velocity is allowed to exceed the Cherenkov threshold. The quantization of moving media has been discussed in Refs. [6-10], but to our knowledge a quantization theory for the case of polarizable bodies moving above the Cherenkov threshold was not developed yet.

Here, we demonstrate that above the Cherenkov threshold the wave energy may become *negative*, such that the wave energy operator has no lower bound. We prove that the hybridization of the guided modes supported by two moving bodies (e.g. such that one of the bodies is at rest in the lab frame and the other moves with a velocity above the Cherenkov threshold) may produce natural modes associated with complex valued frequencies. This phenomenon can occur as long as the velocity of the bodies is kept constant. It is proven that oscillations associated with complex valued frequencies imply an exchange of wave momentum by the moving bodies, and hence in order to keep the velocity of the bodies constant it is necessary to apply an external mechanical force. We suggest that in the framework of a quantum theory, the role of the external force is to counterbalance the effect of the so-called "quantum friction" that recently received significant attention in the scientific literature [7, 11-20].

It should be mentioned that related instabilities of the electromagnetic field have been reported in Ref. [19] for the case of interacting monoatomic layers of coupled electric dipoles, and that the effect of quantum friction was linked to those instabilities. However, the analysis of Ref. [19] is based on a time dependent Hamiltonian, and hence the emergence of system instabilities is less surprising. Here we show that time independent Hamiltonians may also lead to electromagnetic instabilities, and develop a comprehensive theory for the quantization of the macroscopic electromagnetic fields in such systems.

## II. The system under analysis

### A. The wave dynamics

The system in which we are interested in is formed by polarizable moving media, invariant to translations along the *x* and *y*-directions (Fig. 1). The electromagnetic response of each material is described in the respective co-moving frame by the permittivity and permeability functions $\varepsilon = \varepsilon(z)$ and $\mu = \mu(z)$. The velocity of the materials measured with respect to a fixed reference (laboratory) frame is $\mathbf{v} = v(z)\hat{\mathbf{x}}$, being the dependence on *z* a consequence of the fact that we allow different bodies to move with different velocities. In case $\varepsilon$ and $\mu$ are assumed frequency independent, the relativistic relation between the classical **D** and **B** fields and the classical **E** and **H** fields in the laboratory frame is as follows [10, 21]:

$$\begin{pmatrix} \mathbf{D} \\ \mathbf{B} \end{pmatrix} = \begin{pmatrix} \varepsilon_0 \bar{\bar{\varepsilon}} & \frac{1}{c}\bar{\bar{\vartheta}} \\ \frac{1}{c}\bar{\bar{\zeta}} & \mu_0 \bar{\bar{\mu}} \end{pmatrix} \begin{pmatrix} \mathbf{E} \\ \mathbf{H} \end{pmatrix} \equiv \mathbf{M} \cdot \begin{pmatrix} \mathbf{E} \\ \mathbf{H} \end{pmatrix} \tag{1}$$

where the dimensionless parameters $\bar{\bar{\varepsilon}}$, $\bar{\bar{\mu}}$, $\bar{\bar{\vartheta}}$ and $\bar{\bar{\zeta}}$ are such that,

$$\bar{\bar{\varepsilon}} = \varepsilon_t\left(\bar{\bar{\mathbf{I}}} - \hat{\mathbf{x}}\hat{\mathbf{x}}\right) + \varepsilon\hat{\mathbf{x}}\hat{\mathbf{x}} \quad ; \quad \varepsilon_t = \varepsilon\frac{1-\beta^2}{1-n^2\beta^2} \quad (2a)$$

$$\bar{\bar{\mu}} = \mu_t\left(\bar{\bar{\mathbf{I}}} - \hat{\mathbf{x}}\hat{\mathbf{x}}\right) + \mu\hat{\mathbf{x}}\hat{\mathbf{x}} \quad ; \quad \mu_t = \mu\frac{1-\beta^2}{1-n^2\beta^2} \quad (2b)$$

$$\bar{\bar{\zeta}} = -\bar{\bar{\vartheta}} = -a\hat{\mathbf{x}}\times\bar{\bar{\mathbf{I}}} \quad ; \quad a = \beta\frac{n^2-1}{1-n^2\beta^2} \quad (2c)$$

being $\beta = v/c$, and $n^2 = \varepsilon\mu$, $\varepsilon$ and $\mu$ are the material parameters in the rest frame of the pertinent body. The material matrix $\mathbf{M} = \mathbf{M}(z)$ is symmetric and real valued.

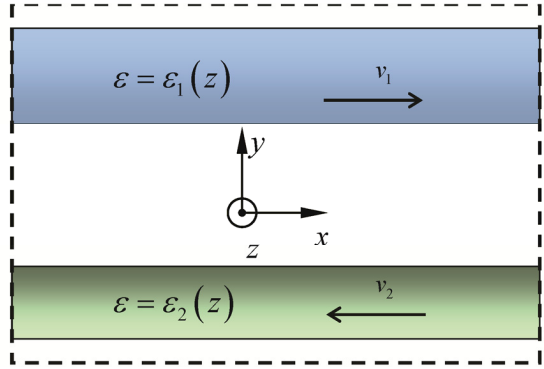

Fig. 1. (Color online) Representative geometry of the system under study: two non-dispersive polarizable bodies move with different velocities with respect to a fixed reference frame. The system is invariant to translations along the $x$-direction.

The electromagnetic field satisfies the Maxwell's equations, which, in the absence of radiation sources, can be written as:

$$\hat{N}\cdot\mathbf{F} = i\mathbf{M}\cdot\frac{\partial \mathbf{F}}{\partial t}, \quad (3)$$

where $\mathbf{F} = \begin{pmatrix} \mathbf{E} \\ \mathbf{H} \end{pmatrix}$, $\hat{N} = \begin{pmatrix} 0 & i\nabla\times \\ -i\nabla\times & 0 \end{pmatrix}$ and $\mathbf{M} = \mathbf{M}(z)$ is the material matrix. The oscillatory states (with a time variation $e^{-i\omega t}$) of the electromagnetic field satisfy:

$$\hat{N} \cdot \mathbf{F}_\omega = \omega \mathbf{M} \cdot \mathbf{F}_\omega. \tag{4}$$

We introduce the sesquilinear form $\langle \ | \ \rangle$ such that for two generic six-component vectors $\mathbf{F}_1$ and $\mathbf{F}_2$ we have:

$$\langle \mathbf{F}_2 | \mathbf{F}_1 \rangle = \frac{1}{2} \int d^3\mathbf{r}\, \mathbf{F}_2^* \cdot \mathbf{M}(\mathbf{r}) \cdot \mathbf{F}_1. \tag{5}$$

In particular, the wave energy can be written in terms of $\langle \ | \ \rangle$ as:

$$H_{EM,P} = \frac{1}{2} \int d^3\mathbf{r}\, \mathbf{B} \cdot \mathbf{H} + \mathbf{D} \cdot \mathbf{E} = \langle \mathbf{F} | \mathbf{F} \rangle. \tag{6}$$

As discussed in our previous work [10], provided the velocity of all the bodies is below the Cherenkov threshold $|v| < c/n$, with $n$ the index of refraction of the moving body in the respective co-moving frame, then $\mathbf{M} = \mathbf{M}(z)$ is positive definite. In such a case, $\langle \ | \ \rangle$ is positive definite and determines an inner product in the space of six-component vectors, and thus the wave energy satisfies $H_{EM,P} \geq 0$ [10]. Moreover, the operator $\mathbf{M}^{-1} \cdot \hat{N}$ is Hermitian,

$$\langle \mathbf{F}_2 | \mathbf{M}^{-1} \cdot \hat{N} \mathbf{F}_1 \rangle = \langle \mathbf{M}^{-1} \cdot \hat{N} \mathbf{F}_2 | \mathbf{F}_1 \rangle, \tag{7}$$

and as a consequence the frequencies associated with the stationary states are real-valued. The spatial-domain is terminated with periodic boundary conditions, and this ensures that the properties of the system are unchanged by a translation along the *x*-direction, i.e. the system is effectively homogeneous for translations along *x* (the slabs are infinitely wide). These properties permit a straightforward quantization of the macroscopic electromagnetic field when $|v_i| < c/n_i$ [10].

## B. Total energy and momentum

Having described the framework adopted to model the macroscopic electrodynamics of the wave fields, next we characterize the mechanical degrees of freedom. As discussed in Ref. [10], the total energy of the system ($H_{tot}$) has a wave part which was already discussed ($H_{EM,P}$), and in addition a part associated with the canonical momentum of the moving bodies,

$$H_{tot} = \sum_i \frac{p_{can,i}^2}{2M_i} + H_{EM,P}, \qquad (8)$$

where $M_i$ is the mass of the $i$-th body, and $p_{can,i}$ is the $x$-component of the total canonical momentum of the $i$-th body (only this component is relevant in the scenarios in which we are interested in). The canonical momentum $\mathbf{p}_{can}$ is the conjugate of the position vector, and it is well-known that for charged particles it differs from the kinetic momentum ($\mathbf{p}_{kin}$) [22, p. 582]. The wave energy [Eq. (6)] may be regarded as the energy stored in the electromagnetic field plus an interaction term which is intrinsically mechanic associated with the "polarization waves", i.e. with the dipole vibrations. Note that Eq. (8) is supposed to hold in the non-relativistic regime so that $v_i/c \ll 1$. Because we are interested in velocities exceeding the Cherenkov threshold ($|v_i| > c/n_i$) it is necessary that $n_i \gg 1$. For continuous media and systems invariant to translations along the $x$-direction, the dynamics of the canonical momentum is determined by [10]:

$$dp_{can,i}/dt = F_{i,x}^{ext}, \qquad (9)$$

where $F_{i,x}^{ext}$ represents an hypothetical external force (of origin not electromagnetic) acting on the $i$-th body. In particular, for a fully closed system ($F_{i,x}^{ext} = 0$) the canonical momentum is preserved, and this can be understood in simple terms as a consequence of $H_{tot}$ being independent of the coordinate $x_{0,i}$ that determines the *x*-position of the center of mass of the *i*-th slab.

It is relevant to find the total force acting on the *i*-th slab. To do this we use the stress tensor theorem [22] which establishes that for the *i*-th body

$$F_{i,x}^L + \frac{dp_{EM,i}}{dt} = \int_{\partial V_i} \hat{\mathbf{n}} \cdot \overline{\overline{\mathbf{T}}} \cdot \hat{\mathbf{x}} \, ds \qquad (10)$$

where $F_{i,x}^L$ the total (*x*-component of the) Lorentz force acting on the pertinent body, $p_{EM,i}$ is the *x*-component of the momentum of the fields in the *i*-th body, $\overline{\overline{\mathbf{T}}}$ is the Maxwell stress-tensor, and $\partial V_i$ is the boundary surface of the body. Noting that the kinetic momentum of the *i-th* body ($p_{kin,i}$) is related to the Lorentz force as $\frac{dp_{kin,i}}{dt} = F_{i,x}^L + F_{i,x}^{ext}$ it is possible to write:

$$\frac{dp_i}{dt} \equiv \frac{dp_{kin,i}}{dt} + \frac{dp_{EM,i}}{dt} = \int_{\partial V_i} \hat{\mathbf{n}} \cdot \overline{\overline{\mathbf{T}}} \cdot \hat{\mathbf{x}} \, ds + F_{i,x}^{ext}. \qquad (11)$$

This equation establishes that the time rate variation of the total momentum $p_i = p_{kin,i} + p_{EM,i}$ associated with the *i-th* body is determined by $F_{i,x}^{ext}$ and by the Maxwell stress tensor. It is crucial to note that the derivation of Eq. (10) is based on a microscopic theory, and hence the fields used in the definition of $\overline{\overline{\mathbf{T}}}$ must be the microscopic electromagnetic fields [22]. However, in case the region exterior to the body is a vacuum the microscopic fields can be identified with the macroscopic fields used in the framework of macroscopic electrodynamics. In these circumstances, Eq. (10) remains valid if $\overline{\overline{\mathbf{T}}}$ is written in terms of the macroscopic fields **E** and **B**. Based on this observation, we prove in Appendix A that for systems invariant to translations along the $x$-direction the stress tensor surface integral can be written in terms of the so-called wave momentum $p_{w,i} = \int_{V_i} (\mathbf{D} \times \mathbf{B}) \cdot \hat{\mathbf{x}} d^3\mathbf{r}$ as:

$$\frac{dp_{kin,i}}{dt} + \frac{dp_{EM,i}}{dt} = \frac{dp_{w,i}}{dt} + F_{i,x}^{ext}. \tag{12}$$

Thus, from Eq. (9) we see that Eq. (12) is consistent with the decompositions for the total momentum $p_i = p_{kin,i} + p_{EM,i} = p_{can,i} + p_{wv,i}$ [10, 23, 24]. In the framework of macroscopic electromagnetism, the electromagnetic momentum is $p_{EM,i} = \frac{1}{c^2} \int_{V_i} (\mathbf{E} \times \mathbf{H}) \cdot \hat{\mathbf{x}} d^3\mathbf{r}$ [23].

Note that the formula for $p_{EM,i}$ (written in terms of the macroscopic fields) does not follow directly from the stress tensor theorem [22], which, as mentioned previously, is derived based on the "microscopic" electromagnetic fields.

In this article, we are interested in the case wherein the velocity of the moving slabs is enforced to be a constant, such that $dp_{kin,i}/dt = 0$. Equation (12) shows that in general this is only possible provided an external force given by,

$$F_{i,x}^{ext} = -\frac{dp_{ps,i}}{dt}, \quad \text{with} \quad p_{ps,i} = \int_{V_i} \left(\mathbf{D} \times \mathbf{B} - \frac{1}{c^2} \mathbf{E} \times \mathbf{H}\right) \cdot \hat{\mathbf{x}} d^3\mathbf{r} \tag{13}$$

is applied to the $i$-*th* slab. In the above, $p_{ps,i} = p_{wv,i} - p_{EM,i} = p_{kin,i} - p_{can,i}$ is the ($x$-component) of the so-called pseudo-momentum of the $i$-th slab [25-26]. For future reference, we also note that from Eqs. (8)-(9) the time-derivative of the total energy is:

$$\begin{aligned}\frac{dH_{tot}}{dt} &= \sum_i \frac{p_{can,i}}{M_i} \frac{dp_{can,i}}{dt} + \frac{dH_{EM,P}}{dt} = \sum_i \frac{p_{can,i}}{M_i} F_{i,x}^{ext} + \frac{dH_{EM,P}}{dt} \\ &\approx \sum_i v_i F_{i,x}^{ext} + \frac{dH_{EM,P}}{dt}\end{aligned}, \tag{14}$$

where the last identity is valid for relatively weak fields and large values of the velocity of the pertinent body.

## III. Instabilities in the limit of weak electromagnetic interaction

### *A. Hybridization of the guided modes*

It is natural to wonder what happens if the velocity of one or more bodies exceeds the Cherenkov threshold so that case the material matrix $\mathbf{M} = \mathbf{M}(z)$ becomes indefinite. It is shown next that, surprisingly, in such circumstances the frequencies of oscillation of the natural modes may become complex-valued. It should be noted at the outset that complex-valued frequencies of oscillation are only allowed if at least two bodies move with different velocities. Otherwise there is a frame where all the bodies are at rest and evidently in such a frame the frequencies of oscillation $\omega$ are real-valued. Hence, the frequencies of oscillation in any other frame can be obtained by a Lorentz transformation of $(\omega, k_x)$ and thus are also real-valued (here $k_x$ is the real-valued wave vector component along *x*).

To illustrate how the electromagnetic coupling of two moving bodies may result in a natural mode of oscillation with a complex valued $\omega$, we consider the problem of interaction of two weakly coupled (e.g. very distant) moving dielectric slabs separated by a vacuum (Fig. 1). Specifically, let us suppose that in the *absence of interaction* each slab supports a guided (trapped) mode described by the electromagnetic field $\mathbf{F}_i = (\mathbf{E}_i \quad \mathbf{H}_i)^T$, where *i*=1,2 identifies the pertinent slab. The modes $\mathbf{F}_1$ and $\mathbf{F}_2$ are associated with the same real valued frequency $\omega'$ and with the same real-valued transverse wave vector $\mathbf{k} = (k_x, k_y, 0)$ [variation along $x$ and $y$ coordinates is of the form $e^{i\mathbf{k}\cdot\mathbf{r}}$] such that in the lab reference frame one has:

$$\hat{N}\cdot\mathbf{F}_1 = \omega'\mathbf{M}_1\cdot\mathbf{F}_1 \qquad \hat{N}\cdot\mathbf{F}_2 = \omega'\mathbf{M}_2\cdot\mathbf{F}_2 \qquad (15)$$

where $\mathbf{M}_1$ represents the material matrix in the scenario wherein the second body is removed, etc. Next, using a perturbation approach we obtain an approximate solution for the eigenvalue problem $\hat{N}\cdot\mathbf{F} = \omega\mathbf{M}\cdot\mathbf{F}$ where $\mathbf{M}$ is the material matrix in the presence of both bodies. To this end, we look for a solution of the form $\mathbf{F} = \alpha_1\mathbf{F}_1 + \alpha_2\mathbf{F}_2$ with unknown coefficients $\alpha_i$ ($i=1,2$). In order that $\mathbf{F} = \alpha_1\mathbf{F}_1 + \alpha_2\mathbf{F}_2$ is a natural mode of oscillation it is necessary that $\alpha_1\omega'\mathbf{M}_1\cdot\mathbf{F}_1 + \alpha_2\omega'\mathbf{M}_2\cdot\mathbf{F}_2 = \omega\mathbf{M}\cdot(\alpha_1\mathbf{F}_1 + \alpha_2\mathbf{F}_2)$. This can also be written as:

$$\alpha_1\omega'(\mathbf{M}_1 - \mathbf{M})\cdot\mathbf{F}_1 + \alpha_2\omega'(\mathbf{M}_2 - \mathbf{M})\cdot\mathbf{F}_2 = \delta\omega\mathbf{M}\cdot(\alpha_1\mathbf{F}_1 + \alpha_2\mathbf{F}_2) \qquad (16)$$

where $\delta\omega = \omega - \omega'$. Calculating the canonical inner product of both sides of the equation with $\mathbf{F}_i$ we obtain the following matrix system:

$$\omega'\begin{pmatrix} \langle \mathbf{F}_1 | \mathbf{M}-\mathbf{M}_1 | \mathbf{F}_1 \rangle_c & \langle \mathbf{F}_1 | \mathbf{M}-\mathbf{M}_2 | \mathbf{F}_2 \rangle_c \\ \langle \mathbf{F}_2 | \mathbf{M}-\mathbf{M}_1 | \mathbf{F}_1 \rangle_c & \langle \mathbf{F}_2 | \mathbf{M}-\mathbf{M}_2 | \mathbf{F}_2 \rangle_c \end{pmatrix}\begin{pmatrix} \alpha_1 \\ \alpha_2 \end{pmatrix} = -\delta\omega\begin{pmatrix} \langle \mathbf{F}_1 | \mathbf{M} | \mathbf{F}_1 \rangle_c & \langle \mathbf{F}_1 | \mathbf{M} | \mathbf{F}_2 \rangle_c \\ \langle \mathbf{F}_2 | \mathbf{M} | \mathbf{F}_1 \rangle_c & \langle \mathbf{F}_2 | \mathbf{M} | \mathbf{F}_2 \rangle_c \end{pmatrix}\begin{pmatrix} \alpha_1 \\ \alpha_2 \end{pmatrix}$$

$$(17)$$

In the above, $\langle \ \rangle_c$ stands for the usual canonical inner product, which should not be confused with the weighted inner product of Eq. (5). Specifically, $\langle \ \rangle_c$ is defined as in Eq. (5) but with $\mathbf{M} = 1$.

Next, we note that in the limit of a weak interaction the anti-diagonal terms in the right-hand side matrix can be neglected as compared to the diagonal terms, because the overlap of $\mathbf{F}_1$ and $\mathbf{F}_2$ is small. Indeed, the fields $\mathbf{F}_i$ decay as $e^{-\gamma_0 d}$ with the distance $d$ away from the *i*-th slab, being $\gamma_0 = \sqrt{k_x^2 + k_y^2 - \omega'^2/c^2}$ the transverse decay constant (along *z*) of the guided mode. Moreover, we can use the approximation $\langle \mathbf{F}_i | \mathbf{M} | \mathbf{F}_i \rangle_c \approx \langle \mathbf{F}_i | \mathbf{M}_i | \mathbf{F}_i \rangle_c \equiv E_{s,i}$ to evaluate the diagonal terms, being $E_{s,i}$ the wave energies associated with the uncoupled guided modes. On the other hand, for the left-hand side matrix the situation is reversed and the diagonal terms are negligible. The reason is that $\mathbf{F}_1 \cdot (\mathbf{M} - \mathbf{M}_1) \cdot \mathbf{F}_1 \sim e^{-2\gamma_0 d}$ and $\mathbf{F}_1 \cdot (\mathbf{M} - \mathbf{M}_1) \cdot \mathbf{F}_2 \sim e^{-\gamma_0 d}$ because both terms are nonzero only over the second slab. This discussion implies that:

$$\begin{pmatrix} 0 & \omega' \langle \mathbf{F}_1 | \mathbf{M} - \mathbf{M}_2 | \mathbf{F}_2 \rangle_c \\ \omega' \langle \mathbf{F}_2 | \mathbf{M} - \mathbf{M}_1 | \mathbf{F}_1 \rangle_c & 0 \end{pmatrix} \begin{pmatrix} \alpha_1 \\ \alpha_2 \end{pmatrix} = -\delta\omega \begin{pmatrix} E_{s,1} & 0 \\ 0 & E_{s,2} \end{pmatrix} \begin{pmatrix} \alpha_1 \\ \alpha_2 \end{pmatrix} \quad (18)$$

Introducing $\Omega_1 = \omega' \langle \mathbf{F}_1 | \mathbf{M} - \mathbf{M}_2 | \mathbf{F}_2 \rangle_c / E_{s,2}$ and $\Omega_2 = \omega' \langle \mathbf{F}_2 | \mathbf{M} - \mathbf{M}_1 | \mathbf{F}_1 \rangle_c / E_{s,1}$, and writing $\alpha_i' = \alpha_i E_{s,i}$, we obtain the following homogeneous system:

$$\begin{pmatrix} \delta\omega & \Omega_1 \\ \Omega_2 & \delta\omega \end{pmatrix} \begin{pmatrix} \alpha_1' \\ \alpha_2' \end{pmatrix} = 0. \quad (19)$$

Thus, the perturbation in the oscillation frequency resulting from the electromagnetic coupling of the guided modes is such that $\delta\omega^2 - \Omega_1 \Omega_2 = 0$, that is:

$$\omega = \omega' \pm \sqrt{\Omega_1 \Omega_2}. \quad (20)$$

It is proven in Appendix B that $\Omega_1 / \Omega_2^* = E_{s,1} / E_{s,2}$. Therefore, we see that depending on the sign of $E_{s,1} / E_{s,2}$ the coupled system is either characterized by *(i)* $E_{s,1} / E_{s,2} > 0$, two natural modes of vibration with real valued frequencies $\omega = \omega' \pm \sqrt{|\Omega_1 \Omega_2|}$, or *(ii)* $E_{s,1} / E_{s,2} < 0$, a pair of natural modes with complex conjugated frequencies $\omega = \omega' \pm i\sqrt{|\Omega_1 \Omega_2|}$. Therefore, when the wave energies $E_{s,1}$ and $E_{s,2}$ associated with the uncoupled guided waves have different signs the hybridization leads to complex-valued frequencies of vibration. This evidently requires that at least one of the slabs moves with velocity above the Cherenkov threshold, because otherwise $\mathbf{M}$ is positive definite and both $E_{s,1}$ and $E_{s,2}$ are positive. From a physical point of view, it is not totally unexpected that the wave energy may become negative for a moving dielectric because (from a quantum perspective) this may be regarded as a consequence of the relativistic relation (relativistic Doppler shift) between wave energy ($\omega$) and wave momentum ($\mathbf{k}$). In Ref. [20], we report rigorous numerical simulations (not based on perturbation theory) that confirm that the hybridization of the guided modes results in system instabilities associated complex-valued frequencies.

To further characterize the response of the system in case of complex-valued frequencies of oscillation, next we study the modal fields. It is evident that the eigenmodes associated with $\omega = \omega' \pm i\lambda$ (denoted by $\mathbf{f}$ and $\mathbf{e}$) with $\lambda = \sqrt{|\Omega_1 \Omega_2|} = |\Omega_1|\sqrt{|E_{s,2}/E_{s,1}|} > 0$ are such that $\begin{pmatrix} \alpha_1 \\ \alpha_2 \end{pmatrix} \sim \begin{pmatrix} -\Omega_1/E_{s,1} \\ \delta\omega/E_{s,2} \end{pmatrix} \sim \begin{pmatrix} \Omega_1/|E_{s,1}| \\ \delta\omega/|E_{s,2}| \end{pmatrix}$, and thus satisfy (apart from an arbitrary normalization factor):

$$\mathbf{f} \sim \frac{\Omega_1}{|E_{s,1}|} \mathbf{F}_1 + \frac{i\lambda}{|E_{s,2}|} \mathbf{F}_2, \qquad \text{for } \omega = \omega' + i\lambda. \tag{21a}$$

$$\mathbf{e} \sim \frac{\Omega_1}{|E_{s,1}|} \mathbf{F}_1 - \frac{i\lambda}{|E_{s,2}|} \mathbf{F}_2, \qquad \text{for } \omega = \omega' - i\lambda. \tag{21b}$$

We fix the normalization factor in such a way that:

$$\mathbf{f} = \frac{\Omega_1}{|\Omega_1|} \frac{1}{\sqrt{2|E_{s,1}|}} \mathbf{F}_1 + \frac{i}{\sqrt{2|E_{s,2}|}} \mathbf{F}_2, \tag{22a}$$

$$\mathbf{e} = \text{sgn}(E_{s,1}) \left( \frac{\Omega_1}{|\Omega_1|} \frac{1}{\sqrt{2|E_{s,1}|}} \mathbf{F}_1 - \frac{i}{\sqrt{2|E_{s,2}|}} \mathbf{F}_2 \right). \tag{22b}$$

where $\text{sgn} = \pm 1$, depending on the sign of the real argument. Note that in the limit of weak coupling the natural mode $\mathbf{f}$ satisfies (below $\langle \ \rangle$ stands for the weighted inner product (5) which yields the wave energy associated with a given mode):

$$\langle \mathbf{f} | \mathbf{f} \rangle = \langle \mathbf{f} | \mathbf{M} | \mathbf{f} \rangle_c \approx \frac{1}{2|E_{s,1}|} \langle \mathbf{F}_1 | \mathbf{M}_1 | \mathbf{F}_1 \rangle_c + \frac{1}{2|E_{s,2}|} \langle \mathbf{F}_2 | \mathbf{M}_2 | \mathbf{F}_2 \rangle_c$$
$$= \frac{E_{s,1}}{2|E_{s,1}|} + \frac{E_{s,2}}{2|E_{s,2}|} = 0 \tag{23a}$$

because $E_{s,1} E_{s,2} < 0$. Based on analogous considerations, it can also be shown that:

$$\langle \mathbf{e} | \mathbf{e} \rangle = 0 \tag{23b}$$

$$\langle \mathbf{e} | \mathbf{f} \rangle = 1 \tag{23c}$$

It is important to stress that the formulas $\langle \mathbf{e} | \mathbf{e} \rangle = \langle \mathbf{f} | \mathbf{f} \rangle = 0$ and $\langle \mathbf{e} | \mathbf{f} \rangle = 1$ *hold at all time instants*. Thus, the wave energy for the modes of oscillation described by $\mathbf{f}$ and $\mathbf{e}$ vanishes, and is independent of time. It should be noted that $\mathbf{f}$ and $\mathbf{e}$ are complex valued fields. The real valued fields $\mathbf{f} + \mathbf{f}^*$ and $\mathbf{e} + \mathbf{e}^*$ have a similar property, e.g. $\langle \mathbf{f} + \mathbf{f}^* | \mathbf{f} + \mathbf{f}^* \rangle = 0$. This is so because $\mathbf{f}$ and $\mathbf{f}^*$ have a spatial variation along $x$ and $y$ of the form $e^{i\mathbf{k}\cdot\mathbf{r}}$ and $e^{-i\mathbf{k}\cdot\mathbf{r}}$, respectively, and this implies that $\langle \mathbf{f} | \mathbf{f}^* \rangle = 0$.

It is striking that $\langle \mathbf{f} | \mathbf{f} \rangle = 0$, i.e. the weighted product of a vector with itself vanishes. This can occur because above the Cherenkov threshold the material matrix $\mathbf{M} = \mathbf{M}(z)$ becomes indefinite, and hence the wave energy may become negative, and $\langle \, | \, \rangle$ is not a positive definite inner product. We shall refer to $\langle \, | \, \rangle$ as an indefinite inner product. Note that even though $\mathbf{f}$ and $\mathbf{e}$ are eigenmodes associated with different frequencies of oscillation they are not orthogonal with respect to the indefinite inner product: $\langle \mathbf{e} | \mathbf{f} \rangle = 1$.

## B. Exchange of energy and momentum and the external force

How is it possible to have $\langle \mathbf{f} | \mathbf{f} \rangle = 0$, i.e. the total wave energy associated with the field $\mathbf{f}$ is conserved and independent of time, even though the field amplitudes vary with time as $e^{-i\omega t} = e^{-i\omega' t} e^{\lambda t}$ (growing exponent)? Let us suppose without loss of generality that $E_{s,1} > 0$ and $E_{s,2} < 0$. Then the decomposition (22a) shows that $\mathbf{f}$ is the superposition of two guided modes with symmetric wave energies and growing amplitudes. Particularly, the energy stored in the slab *1* (*2*) can be identified with the wave energy associated with the field $\frac{\Omega_1}{|\Omega_1|} \frac{1}{\sqrt{2|E_{s,1}|}} \mathbf{F}_1 e^{-i\omega t}$ ($\frac{i}{\sqrt{2|E_{s,2}|}} \mathbf{F}_2 e^{-i\omega t}$) which varies with time as $\frac{1}{2} e^{2\lambda t}$ ($-\frac{1}{2} e^{2\lambda t}$) in normalized unities. Thus, as time passes, the wave energy of the first slab is larger and larger, whereas the wave energy of the second slab is more and more negative, in such a manner that the total wave energy ($H_{EM,P}$) is conserved. Similarly, for the case of the mode $\mathbf{e}$ the energy stored in slabs *1* and *2* vary with time as $\frac{1}{2} e^{-2\lambda t}$ and $-\frac{1}{2} e^{-2\lambda t}$, and thus the amplitudes of the oscillations are progressively weaker and weaker, but the total wave energy is conserved. Thus, we can picture these electromagnetic instabilities as the result of the continuous *exchange of wave energy* by the interacting slabs. The time variation of the field energy attached to the each slab for the two pertinent modes is illustrated in Fig. 2.

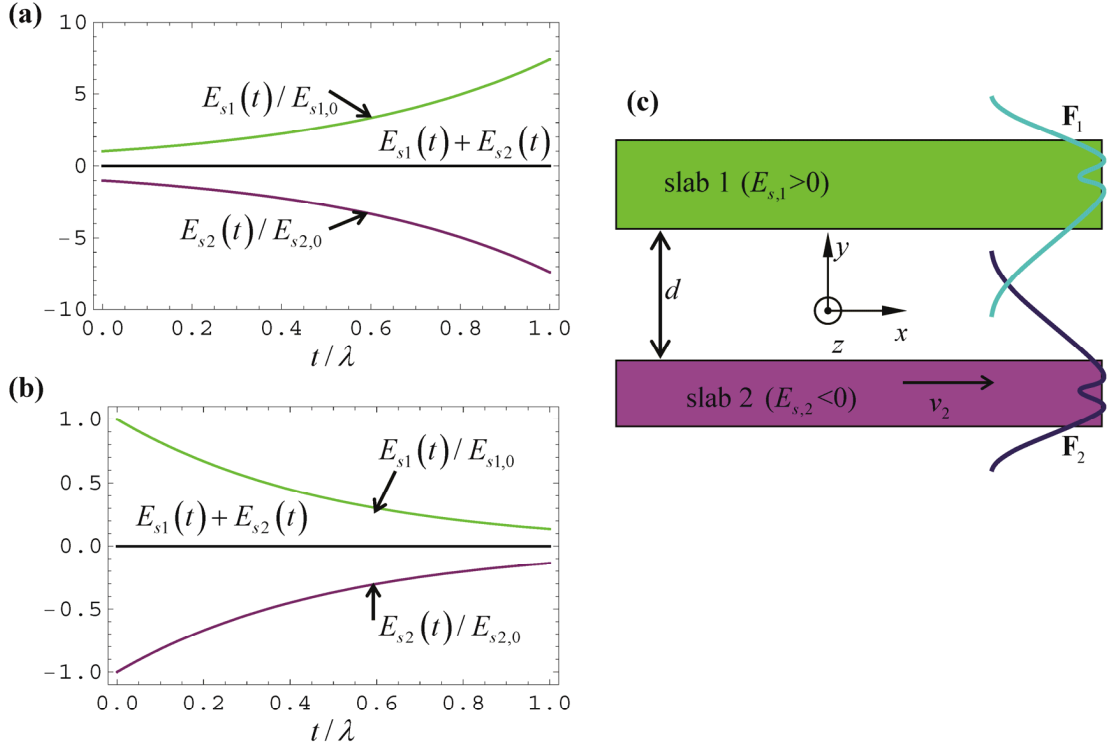

Fig. 2. (Color online) The hybridization of the guided modes $\mathbf{F}_1$ and $\mathbf{F}_2$ defined for the individual slabs results in natural modes of oscillation $\mathbf{f}$ and $\mathbf{e}$ associated with complex conjugated frequencies $\omega = \omega' \pm i\lambda$ (panel c). The wave energy associated with $\mathbf{F}_1$ and $\mathbf{F}_2$ has opposite signs. (a) Normalized wave energy stored in the vicinity of slab 1 ($E_{s1}$) and slab 2 ($E_{s2}$) as a function of time for the mode $\mathbf{f}$ associated with $\omega = \omega' + i\lambda$. (b) Similar to (a) but for the for the mode $\mathbf{e}$ associated with $\omega = \omega' - i\lambda$. Despite the system instability the total wave energy is conserved.

It is natural to ask if the exchange of energy by the two slabs may eventually be accompanied by an exchange of wave momentum. As discussed in Sect. II.B, the wave momentum is associated with the momentum density $(\mathbf{D} \times \mathbf{B}) \cdot \hat{\mathbf{x}}$. From the results Ref.

[10] (Appendix C - Eq. C7) the *x*-component of the wave momentum associated with a given slab (alone in free-space) can be written as $p_{wv} = \frac{k_x}{\omega} \langle \mathbf{F} | \mathbf{F} \rangle$. Hence, the wave momentum stored in the slab *1* (*2*) can be identified with the wave momentum associated with the field $\frac{\Omega_1}{|\Omega_1|} \frac{1}{\sqrt{2|E_{s,1}|}} \mathbf{F}_1 e^{-i\omega t}$ ($\frac{i}{\sqrt{2|E_{s,2}|}} \mathbf{F}_2 e^{-i\omega t}$). Assuming with no loss of generality that $E_{s,1} > 0$ and $E_{s,2} < 0$, it is found that:

$$p_{wv,i} \approx \frac{k_x}{\omega'} E_{s,i}(t). \tag{24}$$

with $E_{s,1}(t) = \frac{1}{2} e^{2\lambda t} = -E_{s,2}(t)$. Thus, similar to the wave energy, the wave momentum of both slabs varies exponentially with time, but in such a manner that the total wave momentum is conserved $p_{wv,1} + p_{wv,2} = 0$. The pseudo-momentum associated with the *i*-th slab can be determined with the help of Eqs. 18 and 21 of Ref. [10] which yield:

$$p_{ps,i} = \left(1 - \beta_i^2\right) \frac{1}{\omega'_i} \frac{\partial \omega'_i}{\partial v_i} E_{s,i}(t). \tag{25}$$

where $v_i$ is the velocity of the *i*-th body, $\beta_i = v_i/c$, and here $\omega'_i = \omega'_i(k_x, k_y, v_i)$ is understood as the dispersion of the guided mode supported by the *i*-th slab in the absence of interaction. It is possible to prove that:

$$\left(1 - \beta_i^2\right) \frac{\partial \omega'_i}{\partial v_i} = k_x - \frac{1}{c^2} \omega'_i \frac{\partial \omega'_i}{\partial k_x} = k_x \left(1 - \frac{v_{ph,i} v_{g,i}}{c^2}\right)$$
$$= k_x \left(1 - \frac{1}{c^2} \frac{v_{ph,i}^{co} + v_i}{1 + v_{ph,i}^{co} v_i / c^2} \frac{v_{g,i}^{co} + v_i}{1 + v_{g,i}^{co} v_i / c^2}\right). \tag{26}$$

where $v_{ph} = \omega/k_x$ and $v_g = \partial\omega/\partial k_x$ are the phase and group velocities in the lab frame, and the $v_{ph}^{co}$ and $v_g^{co}$ are the corresponding parameters calculated in the co-moving frame. Note that $p_{ps,1} + p_{ps,2} = 0$.

Quite importantly, the reported exchange of energy and momentum by the two slabs and the exponentially growing field oscillations can only be sustained if the system is pumped by an external mechanical force that ensures that $p_{kin,i}$ is kept constant. Indeed, even though the "wave part" of our system is closed (in the sense that the total wave energy and wave momentum are conserved), the total system, considering also the degrees of freedom associated with the translational motion of the moving bodies, is not. In fact, using $dH_{EM,P}/dt = 0$ in Eq. (14) it is seen that

$$\frac{dH_{tot}}{dt} = \sum_i v_i F_{i,x}^{ext}, \tag{27}$$

and thus the total energy of the system typically varies with time. From Eq. (13) it is evident that the net power (averaged over time) pumped into the system by the external force vanishes for oscillations with $\lambda = 0$ and is non-zero in presence of wave instabilities ($\lambda \neq 0$). Moreover, for a weak interaction, the total energy stored in the $i$-th body varies with time as

$$\frac{dH_{tot,i}}{dt} = v_i F_{i,x}^{ext} + \frac{dE_{s,i}}{dt} = -v_i \frac{dp_{ps,i}}{dt} + \frac{dE_{s,i}}{dt} = -\left[\frac{v_i}{v_{ph,i}}\left(1 - \frac{v_{ph,i} v_{g,i}}{c^2}\right) - 1\right]\frac{dE_{s,i}}{dt}, \tag{28}$$

where we used Eqs. (13), (25), and (26) and put $v_{ph} = \omega/k_x$ equal to the phase velocity. Let us consider the example of Fig. 2 wherein $E_{s,2} < 0$, $v_1 = 0$, $v_2 > 0$, and suppose that the two slabs are identical and that the guided mode has $k_y = 0$. As argued in the beginning of Sect. III.A, in order that $E_{s,2} < 0$ it is required that $v_2 > c/n$ where $n = n_1 = n_2$ is the refractive index of the two slabs in the respective rest frames. Actually, it is possible to refine the wave instability criterion. Indeed, let us consider the reference frame wherein the velocity of the first slab is $\tilde{v}_1 = -c/n$ (Cherenkov threshold for medium 1) so that the (relativistically transformed) velocity of the second slab is $\tilde{v}_2 = (v_2 - cn^{-1})/(1 - n^{-1}v_2/c)$. To have wave instabilities the material matrix $\tilde{\mathbf{M}}$ associated with slab 2 should be indefinite in this reference frame. This imposes that $|\tilde{v}_2| > c/n$ or equivalently $|v_2 - cn^{-1}|/|1 - n^{-1}v_2/c| > c/n$. Combining this result with $v_2 > c/n$ it is found that $v_2 > \frac{2n}{n^2+1}c$, which is a condition stronger than $v_2 > c/n$. It is proven in Appendix C that $v_2/v_{ph,2}$ is necessarily positive when $E_{s,2} < 0$. Thus, we can state that:

$$\frac{v_2}{v_{ph,2}}\left(1 - \frac{v_{ph,2}v_{g,2}}{c^2}\right) = \frac{v_2}{v_{ph,2}}\left(1 - \frac{1}{c^2}\frac{v_{ph,2}^{co} + v_2}{1 + v_{ph,2}^{co}v_2/c^2}\frac{v_{g,2}^{co} + v_2}{1 + v_{g,2}^{co}v_2/c^2}\right) - 1$$
$$\approx \frac{v_2}{v_{ph,2}}\left(1 - \frac{1}{n^2}\right) - 1 > \frac{2n^2}{n^2+1}\left(1 - \frac{1}{n^2}\right) - 1 = \frac{n^2-3}{n^2+1} > 0 \quad (29)$$

The last inequality follows from $n \gg 1$ which is a condition required in order that our formalism applies (see Sect. II.B). The second identity is obtained using the property $v_{ph,2}^{co}v_2 < 0$ when $E_{s,2} < 0$ [Eq. (C5)], and fact that for waveguides made of dispersionless dielectrics and for $k_y = 0$ one has $|v_{ph,i}^{co}| < c/n$ and $|v_{g,i}^{co}| < c/n$.

Let us consider now the wave associated with the growing field oscillations $\mathbf{f}$. Because $v_1 = 0$ and $E_{s,1} > 0$, it is seen immediately from Eq. (28) that $dH_{tot,1}/dt > 0$. On the other hand, using Eq. (29) we also get $dH_{tot,2}/dt > 0$. Thus, even though $E_{s,2}$ is increasingly negative for this mode of oscillation, the total energy stored in the 2$^{nd}$ body is actually a growing function of time. Hence, despite the local density of wave energy ($H_{EM,P}$) can be negative, the total energy density *is always positive*. This further confirms that the pump of our wave instabilities is the external mechanical force that ensures that the velocity of the moving bodies is invariant, even though this pump does not appear explicitly in Maxwell equations, but only indirectly in the constitutive relations. In the absence of an external force the wave instabilities are pumped by the continuous transfer of kinetic energy of the moving bodies to the electromagnetic field, somewhat analogous to the Cherenkov effect but for uncharged bodies.

Using the theory of this article, we argue in Ref. [20] that the quantum friction force [7, 11-20] is a consequence of the reported system instabilities, and that in the absence of an external action it ultimately prevents the oscillations of the electromagnetic field to grow indefinitely. In the rest of this work, it is assumed that the velocity of the moving bodies is constant, either through the application of an external mechanical force or because the bodies are sufficiently massive so that the change of the velocities in the relevant time window is negligible.

## IV. Natural modes of oscillation above the Cherenkov threshold

The analysis of the previous section assumes a weak interaction between the moving slabs, and is based on perturbation theory. Next, we present a detailed and rigorous study of the natural modes of oscillation of the system that is valid even if the moving slabs are strongly coupled. The analysis is valid for the general case wherein $\mathbf{M} = \mathbf{M}(z)$ may be an indefinite matrix and model two or more moving bodies.

## A. Properties of the eigenfunctions

To characterize the spectrum of $\mathbf{M}^{-1} \cdot \hat{N}$ when $\mathbf{M}$ is indefinite, we start by noting that eigenvectors in different proper subspaces of $\mathbf{M}^{-1} \cdot \hat{N}$ are necessarily linearly independent. Hence, the space generated by the eigenvectors of $\mathbf{M}^{-1} \cdot \hat{N}$ associated with non-zero eigenfrequencies (denoted as $S_{\mathbf{M}^{-1} \cdot \hat{N}}$) can be written as a direct sum of proper subspaces as:

$$S_{\mathbf{M}^{-1} \cdot \hat{N}} = \underbrace{S_{\omega_1} \oplus S_{\omega_2} \oplus ....}_{\substack{\text{Proper subspaces associated with} \\ \text{the real valued eigenvalues } \omega_i \neq 0}} \oplus \underbrace{S_{\omega_{c,1},\omega_{c,1}^*} \oplus S_{\omega_{c,2},\omega_{c,2}^*} \oplus ....}_{\substack{\text{Proper subspaces associated with} \\ \text{the complex valued eigenvalues } \omega_{c,i} \\ \text{such that } \text{Im}\{\omega_{c,i}\} > 0}} . \quad (30)$$

In the above $S_{\omega_i}$ represents the proper-subspace associated with a generic (non-zero) real valued eigenvalue $\omega_i$ of the operator $\mathbf{M}^{-1} \cdot \hat{N}$, and $S_{\omega_{c,i},\omega_{c,i}^*}$ represents the direct sum of the proper-subspaces associated with a generic pair of complex valued eigenvalues, $\omega_{c,i}$ and the respective complex conjugate $\omega_{c,i}^*$, such that $\text{Im}\{\omega_{c,i}\} > 0$.

We note that even though $\langle \, | \, \rangle$ may not be a positive definite inner product the operator $\mathbf{M}^{-1} \cdot \hat{N}$ still satisfies Eq. (7). Let $\mathbf{F}_1$ ($\mathbf{F}_2$) be an eigenvector of $\mathbf{M}^{-1} \cdot \hat{N}$ associated with the eigenfrequency $\omega_1$ ($\omega_2$). Then, it follows that $\omega_1 \langle \mathbf{F}_2 | \mathbf{F}_1 \rangle = \langle \mathbf{F}_2 | \mathbf{M}^{-1} \cdot \hat{N} \mathbf{F}_1 \rangle = \langle \mathbf{M}^{-1} \cdot \hat{N} \mathbf{F}_2 | \mathbf{F}_1 \rangle = \omega_2^* \langle \mathbf{F}_2 | \mathbf{F}_1 \rangle$ and hence

$$(\omega_1 - \omega_2^*)\langle \mathbf{F}_2 | \mathbf{F}_1 \rangle = 0 . \quad (31)$$

This result implies that *two generic eigenvectors* associated with eigenfrequencies $\omega_1 \neq \omega_2^*$ *are orthogonal* with respect to the indefinite inner product $\langle \ | \ \rangle$. Hence, it follows that all the subspaces in the direct sum in Eq. (30) are orthogonal with respect to the inner product $\langle \ | \ \rangle$.

Next, we observe that because $\mathbf{M}$ is real-valued it follows from Eq. (4) and from the definition of $\hat{N}$ that if $\mathbf{F}_\omega$ is an eigenfunction associated with $\omega$ (which we write using the shorthand notation $\mathbf{F}_\omega \leftrightarrow \omega$) then $\mathbf{F}_\omega^*$ is an eigenfunction associated with $-\omega^*$, i.e.

$$\text{if } \mathbf{F}_\omega \leftrightarrow \omega \quad \text{then} \quad \mathbf{F}_\omega^* \leftrightarrow -\omega^*. \tag{32}$$

Furthermore, because our system stays invariant under a rotation of 180º with respect to the *z*-axis then

$$\text{if } \mathbf{F}_\omega \leftrightarrow \omega \quad \text{then} \quad \tilde{\mathbf{F}}_\omega \leftrightarrow \omega^*, \tag{33}$$

where $\tilde{\mathbf{F}}_\omega$ is defined by,

$$\tilde{\mathbf{F}}_\omega(\mathbf{r}) = \begin{pmatrix} \mathbf{R}_{z,\pi} & 0 \\ 0 & -\mathbf{R}_{z,\pi} \end{pmatrix} \cdot \mathbf{F}^*(\mathbf{R}_{z,\pi} \cdot \mathbf{r}) \tag{34}$$

being $\mathbf{R}_{z,\pi} = -(\hat{\mathbf{x}}\hat{\mathbf{x}} + \hat{\mathbf{y}}\hat{\mathbf{y}}) + \hat{\mathbf{z}}\hat{\mathbf{z}}$ the transformation matrix associated with the 180º rotation. In other words, if $\hat{N} \cdot \mathbf{F}_\omega = \omega \mathbf{M} \cdot \mathbf{F}_\omega$ then $\hat{N} \cdot \tilde{\mathbf{F}}_\omega = \omega^* \mathbf{M} \cdot \tilde{\mathbf{F}}_\omega$. Hence, Eqs (32)-(33) demonstrate that if $\omega$ is an eigenvalue of $\mathbf{M}^{-1} \cdot \hat{N}$ then $\pm\omega$ and $\pm\omega^*$ also are. For future reference, we note that explicit calculations show that ($\mathbf{F}_i$ is a generic vector and $\alpha$ is a scalar):

$$\tilde{\tilde{\mathbf{F}}} = \mathbf{F} \tag{35a}$$

$$\widetilde{\alpha \mathbf{F}} = \alpha^* \tilde{\mathbf{F}} \tag{35b}$$

$$\left\langle \mathbf{F}_1 | \tilde{\mathbf{F}}_2 \right\rangle = \left\langle \tilde{\mathbf{F}}_1 | \mathbf{F}_2 \right\rangle^* \tag{35c}$$

where the "~" operator is defined consistently with Eq. (34).

An eigenfunction $\mathbf{F}_\omega$ associated with a complex valued eigenvalue is such that

$$\text{if } \mathbf{F}_\omega \leftrightarrow \omega \text{ with } \omega = \omega' + i\lambda \text{ complex valued, then } \left\langle \mathbf{F}_\omega | \mathbf{F}_\omega \right\rangle = 0. \tag{36}$$

Indeed, if one chooses $\mathbf{F}_2 = \mathbf{F}_1 \equiv \mathbf{F}$ in Eq. (31), it is found that $(\omega - \omega^*)\left\langle \mathbf{F} | \mathbf{F} \right\rangle = 0$, and because $\omega - \omega^* = 2i\lambda \neq 0$ the enunciated result follows. Note that for a positive definite inner product when $\mathbf{F} \neq 0$ we have $\left\langle \mathbf{F} | \mathbf{F} \right\rangle \neq 0$, and in such circumstances the condition $(\omega - \omega^*)\left\langle \mathbf{F} | \mathbf{F} \right\rangle = 0$ would imply that the eigenfrequencies are real-valued. Yet, when $\left\langle \ | \ \right\rangle$ is indefinite it is possible to have $\left\langle \mathbf{F} | \mathbf{F} \right\rangle = 0$ with $\mathbf{F} \neq 0$, and therefore one cannot rule out complex-valued eigenvalues. These results generalize the findings of Sect. III [Eq. 23] to the case of a strong electromagnetic coupling.

## *B. Modal Expansions*

Based on the decomposition (30) and on the property (33), one sees that any six-vector $\mathbf{F}$ lying in the space $S_{\mathbf{M}^{-1} \cdot \hat{N}}$ can be expanded as follows:

$$\mathbf{F} = \sum_{\omega_n \text{ real-valued}} \alpha_n \mathbf{F}_n + \sum_{\lambda_n = \text{Im}\{\omega_{c,n}\} > 0} \left( \beta_n \mathbf{f}_n + \chi_n \mathbf{e}_n \right) \tag{37}$$

where $\mathbf{F}_n$ stand for the elements of a basis of eigenfunctions associated with the real-valued eigenfrequencies $\omega_n \neq 0$, $\mathbf{f}_n$ stand for the elements of a basis of eigenfunctions associated with the complex-valued eigenfrequencies $\omega_{c,n}$ with $\lambda_n = \text{Im}\{\omega_{c,n}\} > 0$, and $\mathbf{e}_n = \tilde{\mathbf{f}}_n$ is the dual of $\mathbf{f}_n$ [defined consistently with Eq. (34)] and is associated with the eigenfrequency $\omega_{c,n}^*$. The parameters $\alpha_n, \beta_n, \chi_n$ are the coefficients of the expansion.

It is relevant to mention that when the inner product $\langle \ | \ \rangle$ is indefinite, the operator $\mathbf{M}^{-1} \cdot \hat{N}$ does not have to be diagonalizable, despite satisfying Eq. (7). In other words, the eigenvectors of $\mathbf{M}^{-1} \cdot \hat{N}$ may not span the whole space. For simplicity it is assumed here that $\mathbf{M}^{-1} \cdot \hat{N}$ is diagonalizable. When $\mathbf{M}^{-1} \cdot \hat{N}$ is not diagonalizable (e.g. if it has a Jordan decomposition) it may be possible to adapt the ideas developed next to obtain a suitable modal expansion and quantization theory, but we will not pursue this here to avoid having a discussion excessively technical.

Under the hypothesis that the eigenfunctions of $\mathbf{M}^{-1} \cdot \hat{N}$ span the whole space, it is proven in Appendix D that it is possible to choose the basis elements $\mathbf{F}_n$ and $\mathbf{f}_n$ such that the following orthogonality conditions hold:

$$\langle \mathbf{e}_n | \mathbf{e}_m \rangle = \langle \mathbf{f}_n | \mathbf{f}_m \rangle = \langle \mathbf{e}_n | \mathbf{F}_m \rangle = \langle \mathbf{f}_n | \mathbf{F}_m \rangle = 0, \tag{38a}$$

$$\langle \mathbf{e}_n | \mathbf{f}_m \rangle = \delta_{n,m}, \tag{38b}$$

$$\langle \mathbf{F}_n | \mathbf{F}_m \rangle = \pm \delta_{n,m}, \qquad \text{$n,m$ arbitrary.} \tag{38c}$$

Thus $S_{\mathbf{M}^{-1} \cdot \hat{N}}$ can be written as a direct sum of orthogonal subspaces with dimension one (associated with real-valued eigenvalues) or two (associated with a pair of complex-conjugate eigenvalues). Note that this result does not follow directly from the decomposition (30) because the proper subspaces can be degenerate. Because the inner product $\langle \ | \ \rangle$ is indefinite, the eigenfunctions $\mathbf{F}_n$ can be associated with either a positive energy (normalization $\langle \mathbf{F}_n | \mathbf{F}_m \rangle = 1$ is adopted) or with a negative energy (normalization $\langle \mathbf{F}_n | \mathbf{F}_n \rangle = -1$ is adopted).

Next, we note that because our system is invariant to translations along the $x$ and $y$ directions, the natural modes depend on $x$ and $y$ as $e^{i\mathbf{k} \cdot \mathbf{r}}$, where $\mathbf{k} = (k_x, k_y, 0)$ is the (real-valued) transverse wave vector. Thus, the eigenfunctions of $\mathbf{M}^{-1} \cdot \hat{N}$ can be assumed to be Bloch waves with a variation of the form $e^{i\mathbf{k} \cdot \mathbf{r}}$ along the transverse coordinates. Clearly, eigenfunctions associated with a different $\mathbf{k}$ are orthogonal. Moreover, from Eqs. (32)-(34) it follows that if $\mathbf{F} \leftrightarrow (\omega, \mathbf{k})$ then $\mathbf{F}^* \leftrightarrow (-\omega^*, -\mathbf{k})$ and $\tilde{\mathbf{F}} \leftrightarrow (\omega^*, \mathbf{k})$. From these properties, we conclude that for a generic *real-valued* field the expansion (37) can be replaced by:

$$\mathbf{F} = \sum_{n\mathbf{k} \in E_R} \left( \alpha_{n\mathbf{k}} \mathbf{F}_{n\mathbf{k}} + \alpha_{n\mathbf{k}}^* \mathbf{F}_{n\mathbf{k}}^* \right) + \sum_{n\mathbf{k} \in E_C} \left( \beta_{n\mathbf{k}} \mathbf{f}_{n\mathbf{k}} + \chi_{n\mathbf{k}} \mathbf{e}_{n\mathbf{k}} + \beta_{n\mathbf{k}}^* \mathbf{f}_{n\mathbf{k}}^* + \chi_{n\mathbf{k}}^* \mathbf{e}_{n\mathbf{k}}^* \right), \tag{39}$$

with $\mathbf{e}_{n\mathbf{k}} = \tilde{\mathbf{f}}_{n\mathbf{k}}$, and with $\mathbf{F}_{n\mathbf{k}} \leftrightarrow \omega_{n\mathbf{k}}$ and $\mathbf{f}_{n\mathbf{k}} \leftrightarrow \omega_{c,n\mathbf{k}} = \omega_{n\mathbf{k}}' + i\lambda_{n\mathbf{k}}$ such that the orthogonality conditions (38) give place to:

$$\langle \mathbf{e}_{n\mathbf{k}} | \mathbf{e}_{m\mathbf{q}} \rangle = \langle \mathbf{f}_{n\mathbf{k}} | \mathbf{f}_{m\mathbf{q}} \rangle = \langle \mathbf{e}_{n\mathbf{k}} | \mathbf{F}_{m\mathbf{q}} \rangle = \langle \mathbf{f}_{n\mathbf{k}} | \mathbf{F}_{m\mathbf{q}} \rangle = 0, \tag{40a}$$

$$\langle \mathbf{e}_{n\mathbf{k}} | \mathbf{f}_{m\mathbf{q}} \rangle = \delta_{n,m} \delta_{\mathbf{k},\mathbf{q}}, \tag{40b}$$

$$\langle \mathbf{F}_{n\mathbf{k}} | \mathbf{F}_{m\mathbf{q}} \rangle = \pm \delta_{n,m} \delta_{\mathbf{k},\mathbf{q}}, \tag{40c}$$

The summations associated with the real-valued (complex-valued) eigenvalues are restricted to the sets $E_R$ ($E_C$) such that:

$$E_R = \left\{ (n, \mathbf{k}) : \omega_{n\mathbf{k}} \langle \mathbf{F}_{n\mathbf{k}} | \mathbf{F}_{n\mathbf{k}} \rangle > 0 \right\}, \tag{41a}$$

$$E_C = \left\{ (n, \mathbf{k}) : \lambda_{n\mathbf{k}} = \text{Im}\{\omega_{c,n\mathbf{k}}\} > 0 \text{ and } k_x > 0 \right\}. \tag{41b}$$

Obviously, the choice of $E_R$ and $E_C$ is not unique. For example, one could as well pick $E_C$ given by $E_C = \{(n,\mathbf{k}): (\lambda_{n\mathbf{k}} > 0 \text{ and } \omega'_{n\mathbf{k}} > 0) \text{ or } (\lambda_{n\mathbf{k}} > 0 \text{ and } \omega'_{n\mathbf{k}} = 0 \text{ and } k_x > 0)\}$, or others. The quantized form of the electromagnetic field operators may depend on the particular choices of $E_R$ and $E_C$, but not the physics. The motivation for our choice of $E_R$ will be clear in the section V.

From the expansion (39) and the orthogonality conditions it is simple to check that the wave energy [Eq. (6)] is given by:

$$H_{EM,P} = \sum_{n\mathbf{k} \in E_R} \mathrm{sgn}(\omega_{n\mathbf{k}})(\alpha_{n\mathbf{k}}\alpha^*_{n\mathbf{k}} + \alpha^*_{n\mathbf{k}}\alpha_{n\mathbf{k}})$$
$$+ \sum_{n\mathbf{k} \in E_C} (\chi^*_{n\mathbf{k}}\beta_{n\mathbf{k}} + \beta^*_{n\mathbf{k}}\chi_{n\mathbf{k}} + \beta_{n\mathbf{k}}\chi^*_{n\mathbf{k}} + \chi_{n\mathbf{k}}\beta^*_{n\mathbf{k}}) \tag{42}$$

where $\mathrm{sgn}(.) = \pm 1$ depending on the sign of the argument. Note that from Eq. (41a) $\mathrm{sgn}(\omega_{n\mathbf{k}}) = \mathrm{sgn}(\langle \mathbf{F}_{n\mathbf{k}} | \mathbf{F}_{n\mathbf{k}} \rangle)$. Formula (42) makes manifest that the wave energy is an indefinite quadratic form of the expansion coefficients.

## V. Quantization of the system

In this section, we develop the theory of quantization of the electromagnetic field for a system with moving bodies allowing velocities above the Cherenkov threshold. Similar to our previous work [10], only the wave part of the energy is quantized. The parcel of the energy associated with the canonical momentum can be treated semi-classically [10].

## A. The Hamiltonian

In order to generalize the usual field quantization process to the system under study, we start by promoting $\alpha_{n\mathbf{k}}, \beta_{n\mathbf{k}}, \chi_{n\mathbf{k}}$ and $\alpha_{n\mathbf{k}}^*, \beta_{n\mathbf{k}}^*, \chi_{n\mathbf{k}}^*$ to the operators $\hat{\alpha}_{n\mathbf{k}}, \hat{\beta}_{n\mathbf{k}}, \hat{\chi}_{n\mathbf{k}}$ and $\hat{\alpha}_{n\mathbf{k}}^\dagger, \hat{\beta}_{n\mathbf{k}}^\dagger, \hat{\chi}_{n\mathbf{k}}^\dagger$, respectively. In this manner we readily obtain the Hamiltonian:

$$\hat{H}_{EM,P} = \sum_{n\mathbf{k} \in E_R} \text{sgn}(\omega_{n\mathbf{k}}) \left( \hat{\alpha}_{n\mathbf{k}} \hat{\alpha}_{n\mathbf{k}}^\dagger + \hat{\alpha}_{n\mathbf{k}}^\dagger \hat{\alpha}_{n\mathbf{k}} \right) + \sum_{n\mathbf{k} \in E_C} \left( \hat{\chi}_{n\mathbf{k}}^\dagger \hat{\beta}_{n\mathbf{k}} + \hat{\beta}_{n\mathbf{k}}^\dagger \hat{\chi}_{n\mathbf{k}} + \hat{\beta}_{n\mathbf{k}} \hat{\chi}_{n\mathbf{k}}^\dagger + \hat{\chi}_{n\mathbf{k}} \hat{\beta}_{n\mathbf{k}}^\dagger \right) . \tag{43}$$

To find the commutation relations satisfied by the relevant operators we impose that for a generic operator $\hat{A}$ (which may represent either $\hat{\alpha}_{n\mathbf{k}}, \hat{\beta}_{n\mathbf{k}}, \hat{\chi}_{n\mathbf{k}}$), the time derivative $\dfrac{d\hat{A}}{dt} = \dfrac{i}{\hbar}\left[\hat{H}, \hat{A}\right]$ agrees with the time derivative calculated classically. For example, classically $\dfrac{d\alpha_{n\mathbf{k}}}{dt} = -i\omega_{n\mathbf{k}} \alpha_{n\mathbf{k}}$ and hence we impose that $\dfrac{i}{\hbar}\left[\hat{H}_{EM,P}, \hat{\alpha}_{n\mathbf{k}}\right] = -i\omega_{n\mathbf{k}} \hat{\alpha}_{n\mathbf{k}}$. This procedure yields:

$$\left[\hat{H}_{EM,P}, \hat{\alpha}_{n\mathbf{k}}\right] = -\hbar \omega_{n\mathbf{k}} \hat{\alpha}_{n\mathbf{k}} , \tag{44a}$$

$$\left[\hat{H}_{EM,P}, \hat{\beta}_{n\mathbf{k}}\right] = -\hbar \omega_{c,n\mathbf{k}} \hat{\beta}_{n\mathbf{k}} , \quad \left[\hat{H}_{EM,P}, \hat{\chi}_{n\mathbf{k}}\right] = -\hbar \omega_{c,n\mathbf{k}}^* \hat{\chi}_{n\mathbf{k}} . \tag{44b}$$

We impose that the commutator of any two operators of the type $\hat{\alpha}, \hat{\beta}, \hat{\chi}$ is a scalar. In addition, for simplicity, it is assumed at the outset that operators associated with different indices $n\mathbf{k}$ commute, and that the operators $\hat{\alpha}$ commute with the operators $\hat{\beta}, \hat{\chi}$. Within these hypotheses, it is simple to check that the commutation relations (44) are satisfied provided (we omit the subscripts $n\mathbf{k}$ in some of the formulas below):

$$\text{sgn}(\omega)\left[\hat{\alpha}, \hat{\alpha}^\dagger\right]\hat{\alpha} = \frac{\hbar}{2}\omega\hat{\alpha} , \tag{45a}$$

$$\left[\hat{\chi}^{\dagger},\hat{\beta}\right]\hat{\beta}+\left[\hat{\chi},\hat{\beta}\right]\hat{\beta}^{\dagger}+\left[\hat{\beta}^{\dagger},\hat{\beta}\right]\hat{\chi}=-\frac{\hbar}{2}\omega_{c}\hat{\beta},\qquad(45b)$$

$$\left[\hat{\beta}^{\dagger},\hat{\chi}\right]\hat{\chi}+\left[\hat{\beta},\hat{\chi}\right]\hat{\chi}^{\dagger}+\left[\hat{\chi}^{\dagger},\hat{\chi}\right]\hat{\beta}=-\frac{\hbar}{2}\omega_{c}^{*}\hat{\chi}.\qquad(45c)$$

Hence, from Eq. (45a) it readily follows that,

$$\hat{\alpha}=\sqrt{\frac{\hbar|\omega|}{2}}\hat{c},\qquad \text{for some } \hat{c} \text{ such that } \left[\hat{c},\hat{c}^{\dagger}\right]=1.\qquad(46)$$

On the other hand, calculating the commutator of both members of Eq. (45b) [Eq. (45c)] with $\hat{\beta}$ [$\hat{\chi}$] one finds that $\left[\hat{\beta}^{\dagger},\hat{\beta}\right]\left[\hat{\chi},\hat{\beta}\right]=0$ and $\left[\hat{\chi}^{\dagger},\hat{\chi}\right]\left[\hat{\chi},\hat{\beta}\right]=0$, respectively. Hence, at least two out of the three commutators $\left[\hat{\beta}^{\dagger},\hat{\beta}\right]$, $\left[\hat{\chi}^{\dagger},\hat{\chi}\right]$, and $\left[\hat{\chi},\hat{\beta}\right]$ must vanish. A detailed analysis shows that in order that Eqs. (45b) and (45c) are both satisfied it is necessary that the three commutators vanish. Hence, from (45b) it also follows that $\left[\hat{\chi}^{\dagger},\hat{\beta}\right]=-\frac{\hbar}{2}\omega_{c}$. Moreover, because $\left[\hat{\chi}^{\dagger},\hat{\beta}\right]$ is a scalar we have $\left[\hat{\beta}^{\dagger},\hat{\chi}\right]=\left[\hat{\chi}^{\dagger},\hat{\beta}\right]^{\dagger}=\left[\hat{\chi}^{\dagger},\hat{\beta}\right]^{*}=-\frac{\hbar}{2}\omega_{c}^{*}$, and hence Eq. (45c) is also satisfied. Thus, we demonstrated that the required commutation relations for $\hat{\beta}$ and $\hat{\chi}$ are:

$$\left[\hat{\beta}^{\dagger},\hat{\beta}\right]=\left[\hat{\chi}^{\dagger},\hat{\chi}\right]=\left[\hat{\chi},\hat{\beta}\right]=0\qquad(47a)$$

$$\left[\hat{\chi}^{\dagger},\hat{\beta}\right]=-\frac{\hbar}{2}\omega_{c}\qquad(47b)$$

It can be checked that the $\hat{\beta}$ and $\hat{\chi}$ operators can be written in terms of creation and annihilation operators satisfying standard commutation relations as follows:

$$\hat{\beta}=\frac{1}{2}\sqrt{\hbar\omega_{c}}\left(\hat{a}_{c}+\hat{b}_{c}^{\dagger}\right),\qquad \hat{\chi}=\frac{1}{2}\sqrt{\hbar\omega_{c}^{*}}\left(\hat{a}_{c}-\hat{b}_{c}^{\dagger}\right)\qquad(48a)$$

$$\left[\hat{a}_c,\hat{a}_c^\dagger\right]=\left[\hat{b}_c,\hat{b}_c^\dagger\right]=1, \qquad \left[\hat{a}_c,\hat{b}_c\right]=\left[\hat{a}_c,\hat{b}_c^\dagger\right]=0 \tag{48b}$$

We note that for $\omega_c = \omega' + i\lambda$ the above relations imply that:

$$\hat{\chi}^\dagger\hat{\beta}+\hat{\beta}^\dagger\hat{\chi}+\hat{\beta}\hat{\chi}^\dagger+\hat{\chi}\hat{\beta}^\dagger = \frac{\hbar}{2}\omega'\left(\hat{a}_c\hat{a}_c^\dagger+\hat{a}_c^\dagger\hat{a}_c-\hat{b}_c\hat{b}_c^\dagger-\hat{b}_c^\dagger\hat{b}_c\right)+i\hbar\lambda\left(\hat{a}_c^\dagger\hat{b}_c^\dagger-\hat{a}_c\hat{b}_c\right). \tag{49}$$

Thus, a pair of modes associated with complex-conjugated frequencies is described by two coupled quantum harmonic oscillators associated with symmetric frequencies $\omega'$ and $-\omega'$. Note that the interaction term $i\hbar\lambda\left(\hat{a}_c^\dagger\hat{b}_c^\dagger-\hat{a}_c\hat{b}_c\right)$ does not preserve the occupation numbers.

Feeding Eqs. (46) and (48) into Eq. (43) and restoring the subscripts $n\mathbf{k}$, we obtain the following formula for the quantized Hamiltonian of the system:

$$\hat{H}_{EM,P} = \hat{H}_R + \hat{H}_C \tag{50a}$$

$$\hat{H}_R = \sum_{n\mathbf{k}\in E_R} \frac{\hbar\omega_{n\mathbf{k}}}{2}\left(\hat{c}_{n\mathbf{k}}\hat{c}_{n\mathbf{k}}^\dagger+\hat{c}_{n\mathbf{k}}^\dagger\hat{c}_{n\mathbf{k}}\right) \tag{50b}$$

$$\begin{aligned}\hat{H}_C &= \sum_{n\mathbf{k}\in E_C}\left(\hat{\chi}_{n\mathbf{k}}^\dagger\hat{\beta}_{n\mathbf{k}}+\hat{\beta}_{n\mathbf{k}}^\dagger\hat{\chi}_{n\mathbf{k}}+\hat{\beta}_{n\mathbf{k}}\hat{\chi}_{n\mathbf{k}}^\dagger+\hat{\chi}_{n\mathbf{k}}\hat{\beta}_{n\mathbf{k}}^\dagger\right) \\ &= \sum_{n\mathbf{k}\in E_C}\left[\frac{\hbar\omega'_{n\mathbf{k}}}{2}\left(\hat{a}_{c,n\mathbf{k}}\hat{a}_{c,n\mathbf{k}}^\dagger+\hat{a}_{c,n\mathbf{k}}^\dagger\hat{a}_{c,n\mathbf{k}}-\hat{b}_{c,n\mathbf{k}}\hat{b}_{c,n\mathbf{k}}^\dagger-\hat{b}_{c,n\mathbf{k}}^\dagger\hat{b}_{c,n\mathbf{k}}\right) \right. \\ &\qquad \left. +i\hbar\lambda_{n\mathbf{k}}\left(\hat{a}_{c,n\mathbf{k}}^\dagger\hat{b}_{c,n\mathbf{k}}^\dagger-\hat{a}_{c,n\mathbf{k}}\hat{b}_{c,n\mathbf{k}}\right)\right]\end{aligned} \tag{50c}$$

where $\hat{\beta}_{n\mathbf{k}},\hat{\chi}_{n\mathbf{k}}$ are defined consistently with Eq. (48), and the operators $\hat{a}_{c,n\mathbf{k}},\hat{b}_{c,n\mathbf{k}},\hat{c}_{n\mathbf{k}}$ satisfy standard commutation relations for creation and annihilation operators:

$$\left[\hat{a}_{c,m\mathbf{q}},\hat{a}_{c,n\mathbf{k}}^\dagger\right]=\left[\hat{b}_{c,m\mathbf{q}},\hat{b}_{c,n\mathbf{k}}^\dagger\right]=\left[\hat{c}_{m\mathbf{q}},\hat{c}_{n\mathbf{k}}^\dagger\right]=\delta_{m\mathbf{q},n\mathbf{k}}, \tag{51a}$$

$$\begin{aligned}\left[\hat{a}_{c,m\mathbf{q}},\hat{b}_{c,n\mathbf{k}}\right]=\left[\hat{a}_{c,m\mathbf{q}},\hat{b}_{c,n\mathbf{k}}^\dagger\right]=\left[\hat{a}_{c,m\mathbf{q}},\hat{c}_{n\mathbf{k}}\right]= \\ \left[\hat{a}_{c,m\mathbf{q}},\hat{c}_{n\mathbf{k}}^\dagger\right]=\left[\hat{b}_{c,m\mathbf{q}},\hat{c}_{n\mathbf{k}}\right]=\left[\hat{b}_{c,m\mathbf{q}},\hat{c}_{n\mathbf{k}}^\dagger\right]=0\end{aligned}. \tag{51b}$$

Because $\left[\hat{H}_R, \hat{H}_C\right] = 0$, the eigenstates of $\hat{H}_{EM,P}$ are determined by the direct product of the eigenstates of $\hat{H}_R$ and $\hat{H}_C$. Let us note that $\hat{H}_{EM,P}$ is a generalization of the quantized Hamiltonian for systems such that the inner product is positive (e.g. see [10]). When the inner product is positive definite the set $E_C$ is empty, and thus $\hat{H}_{EM,P} = \hat{H}_R$. Moreover, because all the natural modes can be chosen such that $\langle \mathbf{F}_{n\mathbf{k}} | \mathbf{F}_{n\mathbf{k}} \rangle = 1$ it follows that $E_R$ reduces to $E_R = \{(n,\mathbf{k}): \omega_{n\mathbf{k}} > 0\}$, i.e. all the eigenfrequencies are positive. Thus, in the usual case, the eigenvalues of the energy operator are of the form $\hbar\omega_{n\mathbf{k}}\left(m_{n\mathbf{k}} + \frac{1}{2}\right) + ... + \hbar\omega_{n'\mathbf{k}'}\left(m'_{n'\mathbf{k}'} + \frac{1}{2}\right) > 0$, where $m$ and $m'$ represent nonnegative integers. Quite differently, in case of an indefinite inner product, the basis of the eigenfunctions typically must contain modes normalized as $\langle \mathbf{F}_{n\mathbf{k}} | \mathbf{F}_{n\mathbf{k}} \rangle = -1$, and as a consequence, if one wishes to associate the corresponding quantum oscillators with creation and annihilation operators that satisfy the usual commutation relations, it is necessary to pick the negative eigenfrequencies $\omega_{n\mathbf{k}}$ to be in the set $E_R$. This is the motivation for defining $E_R$ as in Eq. (41). Because of this the spectrum of $\hat{H}_R$ is still of the form $\hbar\omega_{n\mathbf{k}}\left(m_{n\mathbf{k}} + \frac{1}{2}\right) + ... + \hbar\omega_{n'\mathbf{k}'}\left(m'_{n'\mathbf{k}'} + \frac{1}{2}\right)$, but now the eigenvalues can be negative. The energy spectrum of $\hat{H}_{EM,P} = \hat{H}_R + \hat{H}_C$ is analyzed in detail in Sect. V.C.

## B. The quantized macroscopic fields

The quantized electromagnetic field is obtained by replacing $\alpha_{n\mathbf{k}}, \beta_{n\mathbf{k}}, \chi_{n\mathbf{k}}$ in Eq. (39) by the corresponding quantum operators. In the Heisenberg picture the quantized fields are given by:

$$\hat{\mathbf{F}} = \begin{pmatrix} \hat{\mathbf{E}} \\ \hat{\mathbf{H}} \end{pmatrix} = \hat{\mathbf{F}}_R + \hat{\mathbf{F}}_C \tag{52a}$$

$$\hat{\mathbf{F}}_R = \sum_{n\mathbf{k} \in E_R} \sqrt{\frac{\hbar |\omega_{n\mathbf{k}}|}{2}} \left( \hat{c}_{n\mathbf{k}} e^{-i\omega_{n\mathbf{k}} t} \mathbf{F}_{n\mathbf{k}} + \hat{c}^\dagger_{n\mathbf{k}} e^{+i\omega_{n\mathbf{k}} t} \mathbf{F}^*_{n\mathbf{k}} \right) \tag{52b}$$

$$\hat{\mathbf{F}}_C = \sum_{n\mathbf{k} \in E_C} \left( \hat{\beta}_{n\mathbf{k}} e^{-i\omega_{c,n\mathbf{k}} t} \mathbf{f}_{n\mathbf{k}} + \hat{\chi}_{n\mathbf{k}} e^{-i\omega^*_{c,n\mathbf{k}} t} \mathbf{e}_{n\mathbf{k}} + \hat{\beta}^\dagger_{n\mathbf{k}} e^{i\omega^*_{c,n\mathbf{k}} t} \mathbf{f}^*_{n\mathbf{k}} + \hat{\chi}^\dagger_{n\mathbf{k}} e^{i\omega_{c,n\mathbf{k}} t} \mathbf{e}^*_{n\mathbf{k}} \right) \tag{52c}$$

The quantized $\hat{\mathbf{D}}$ and $\hat{\mathbf{B}}$ fields are defined by,

$$\begin{pmatrix} \hat{\mathbf{D}} \\ \hat{\mathbf{B}} \end{pmatrix} \equiv \hat{\mathbf{G}} = \mathbf{M} \cdot \hat{\mathbf{F}} \tag{53}$$

where $\mathbf{M}$ is the material matrix. We prove in Appendix E that provided the set of eigenfunctions of $\mathbf{M}^{-1} \cdot \hat{N}$ spans the considered space of six-vectors then the following equal-time commutation relations are satisfied:

$$\left[ \hat{\mathbf{G}}(\mathbf{r},t), \hat{\mathbf{G}}(\mathbf{r}',t) \right] = \hbar \hat{N} \cdot \left\{ \mathbf{I}_{6\times 6} \delta(\mathbf{r}-\mathbf{r}') \right\}. \tag{54}$$

In the above, $\left[ \hat{\mathbf{G}}(\mathbf{r}), \hat{\mathbf{G}}(\mathbf{r}') \right]$ should be understood as a tensor (with elements $\left[ \hat{G}_m(\mathbf{r}), \hat{G}_n(\mathbf{r}') \right]$, $m,n=1,\ldots 6$,), and $\mathbf{I}_{6\times 6}$ represents the identity tensor. In particular, this result implies that:

$$\left[ \hat{\mathbf{D}}(\mathbf{r},t), \hat{\mathbf{B}}(\mathbf{r}',t) \right] = i\hbar \nabla \times \left\{ \mathbf{I}_{3\times 3} \delta(\mathbf{r}-\mathbf{r}') \right\}, \tag{55a}$$

$$\left[ \hat{\mathbf{D}}(\mathbf{r},t), \hat{\mathbf{D}}(\mathbf{r}',t) \right] = \left[ \hat{\mathbf{B}}(\mathbf{r},t), \hat{\mathbf{B}}(\mathbf{r}',t) \right] = 0. \tag{55b}$$

As already discussed in Ref. [10], the commutation relations for the macroscopic fields are not coincident with the commutation relations in vacuum for the microscopic fields. The reader is referred to Ref. [10] for more details.

It is obvious that because of Eq. (44) the field operators in the Heisenberg picture satisfy the Maxwell's Equations (see Eq. (3)):

$$i\partial_t \hat{\mathbf{G}} = \hat{N}\hat{\mathbf{F}}. \tag{56}$$

Moreover, the Hamiltonian (43) [see also Eq. (50)] can be written in terms of the quantized fields as follows (compare with Eq. (6)):

$$\hat{H}_{EM,P} = \frac{1}{2}\int d^3\mathbf{r}\, \hat{\mathbf{B}}\cdot\hat{\mathbf{H}} + \hat{\mathbf{D}}\cdot\hat{\mathbf{E}} = \frac{1}{2}\int d^3\mathbf{r}\, \hat{\mathbf{F}}\cdot\mathbf{M}\cdot\hat{\mathbf{F}}. \tag{57}$$

These results demonstrate that the quantization is independent of the specific choice of the set $E_C$, because for other choices (see Eq. (41) and the associated discussion) one obtains exactly the same commutation relations for the field operators.

## C. The energy spectrum

As mentioned in Sect. V.A, the eigenstates of $\hat{H}_{EM,P} = \hat{H}_R + \hat{H}_C$ are the tensor product of the eigenstates of $\hat{H}_R$ and $\hat{H}_C$.

The eigenstates of $\hat{H}_R$ can be represented in terms of the occupation numbers of the quantum oscillators $n\mathbf{k} \in E_R$, and thus are denoted by $|..., m_{n\mathbf{k}},..., m'_{n'\mathbf{k}'},...\rangle$ such that the entry $m_{n\mathbf{k}}$ indicates that there are $m$ quanta associated with the oscillator with label $n\mathbf{k}$.

The corresponding energy is $E = ... + \hbar\omega_{n\mathbf{k}}\left(m_{n\mathbf{k}} + \frac{1}{2}\right) + ... + \hbar\omega_{n'\mathbf{k}'}\left(m'_{n'\mathbf{k}'} + \frac{1}{2}\right) + ...$. For an indefinite inner product $\omega_{n\mathbf{k}}$ may be negative, and hence it should be clear that in such a case the wave energy has no lower bound, and in particular $\hat{H}_R$ does not have a ground state. More specifically, for oscillators with $\omega_{n\mathbf{k}}$ negative, the corresponding state $|0_{n\mathbf{k}}\rangle$ is the state of highest energy ($\frac{\hbar}{2}\omega_{n\mathbf{k}} < 0$), and increasing quantum numbers give rise to more negative values of the energy. Therefore, for oscillators with $\omega_{n\mathbf{k}}$ negative, the ladder operators are such that $\hat{c}_{n\mathbf{k}}$ originates states with higher (less negative) wave energy and $\hat{c}^\dagger_{n\mathbf{k}}$ creates states with lower (more negative) energy. This is illustrated in Fig. 3, and contrasted with the standard case wherein $\omega_{n\mathbf{k}}$ is positive.

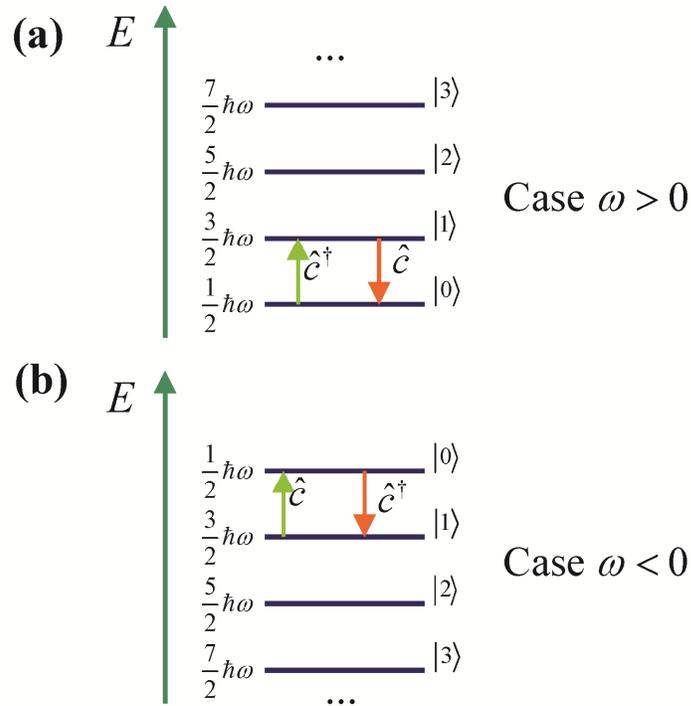

Fig. 3. (Color online) The behavior of the ladder operators is reversed in case of quantum oscillators with a negative $\omega$ (panel *b*) as compared to the standard case where $\omega$ is positive (panel *a*).

What is the meaning of an oscillator with negative energy? Making an analogy with a mechanical system, one can imagine such an oscillator as being formed by a body with *negative* mass subject to the driving force of an elastic spring with a *negative* force constant. The peculiar thing is that the larger is the amplitude of the oscillations the lower is the energy of the system. Thus, when $\omega<0$ the state associated with minimal oscillation amplitude corresponds to the state of highest energy. Because the quanta of $\hat{H}_R$ are bosons, states with negative energies are always vacant. Thus, in some sense, an oscillator with negative frequency can be regarded as never ending reservoir of energy that can be used to pump the oscillators with positive frequency. As discussed previously, oscillators with a negative frequency can only be present if the velocity of the polarizable dielectrics is sufficiently large and is kept constant through the application of an external force that counterbalances the frictional force associated with the radiation drag by the moving bodies [20].

Let us now analyze the spectrum of $\hat{H}_C$. Because $\hat{H}_C$ is written in terms of standard creation and annihilation operators ($\hat{a}_{c,n\mathbf{k}}, \hat{b}_{c,n\mathbf{k}}$, see Eq. 50c), a generic state of $\hat{H}_C$ is a superposition of states of the type $|C\rangle = |...,(m^a_{c,n\mathbf{k}}, m^b_{c,n\mathbf{k}}),...,(m'^a_{c,n'\mathbf{k}'}, m'^b_{c,n'\mathbf{k}'}),...\rangle$, where $(m^a_{c,n\mathbf{k}}, m^b_{c,n\mathbf{k}})$ are the occupation numbers of the oscillators associated with $\hat{a}_{c,n\mathbf{k}}, \hat{b}_{c,n\mathbf{k}}$. However, it is crucial to note that a ket of type $|C\rangle$ *is not* an eigenstate of $\hat{H}_C$. This is so because the terms $\hat{a}^\dagger_{c,n\mathbf{k}}\hat{b}^\dagger_{c,n\mathbf{k}}$ and $\hat{a}_{c,n\mathbf{k}}\hat{b}_{c,n\mathbf{k}}$ in Eq. (50c) do not preserve the occupation numbers. Moreover, still from Eq. (50c), it is seen that $\langle C|\hat{H}_C|C\rangle = \sum_{n\mathbf{k}\in E_C} \hbar\omega'_{n\mathbf{k}} \left(m^a_{c,n\mathbf{k}} - m^b_{c,n\mathbf{k}}\right)$ hence it should be clear that depending on $|C\rangle$, $\langle C|\hat{H}_C|C\rangle$ may be either positive or negative, and thus $\hat{H}_C$ is not a positive definite operator.

To determine the spectrum of $\hat{H}_C$ it is enough to consider the case wherein there is a single pair of "complex oscillators" described by the operators $\hat{a}_c, \hat{b}_c$, or equivalently by $\hat{\beta}$ and $\hat{\chi}$ [Eq. (48)] (the label $n\mathbf{k}$ is dropped in the following). Let us suppose that $|E\rangle$ is some hypothetical eigenstate of $\hat{H}_C$, such that $\hat{H}_C|E\rangle = E|E\rangle$. Then, using the commutation relations (44) it follows that:

$$\begin{aligned}\hat{H}_C\hat{\beta}|E\rangle &= \left[\hat{H}_C, \hat{\beta}\right]|E\rangle + \hat{\beta}\hat{H}_C|E\rangle \\ &= (E - \hbar\omega_c)\hat{\beta}|E\rangle\end{aligned} \quad (58)$$

This proves that $\hat{\beta}|E\rangle = 0$ because otherwise $\hat{\beta}|E\rangle$ would be an eigenstate of $\hat{H}_C$ associated with a complex-valued eigenvalue ($E - \hbar\omega_c$), which is impossible because $\hat{H}_C$ is Hermitian. Using similar arguments, one can prove that $\chi^\dagger|E\rangle = 0$. Thus, this implies that $\left[\hat{\chi}^\dagger, \hat{\beta}\right]|E\rangle = 0$. But from Eq. (47b), one has the commutation relation $\left[\hat{\chi}^\dagger, \hat{\beta}\right] = -\frac{\hbar}{2}\omega_c$, and hence $\left[\hat{\chi}^\dagger, \hat{\beta}\right]|E\rangle = 0$ is equivalent to $|E\rangle = 0$. Hence, we demonstrated that the Hermitian operator $\hat{H}_C$ *does not have eigenfunctions*, or in other words $\hat{H}_C$ does not have stationary states! Because $\hat{H}_{EM,P} = \hat{H}_R + \hat{H}_C$, it follows that when $\hat{H}_C \neq 0$ the total Hamiltonian has the same property.

At first, this may look surprising and unsettling, but actually, from a physical point of view, it is fully consistent with the fact that our system may be unstable above the Cherenkov threshold, and hence it cannot have any state that remains qualitatively the same, as time passes. This is manifested in the time dynamics of the operators $\hat{\beta}$ and $\hat{\chi}$, which is such that the operators can grow or decay exponentially in time, e.g. $\hat{\beta}(t) = \hat{\beta}(t=0)e^{-i\omega_c t}$ with $\omega_c = \omega' + i\lambda$.

From a mathematical point of view, it may look peculiar that $\hat{H}_C$ is Hermitian and does not have eigenfunctions. Indeed, one is quite used to the idea that the spectrum of Hermitian operators spans the entire functional space. However, the spectral theorem usually applies to self-adjoint bounded operators. For example, in the usual situation wherein $\hat{H}$ is positive its spectrum can be assumed to have a positive lower bound, and hence the spectrum of $\hat{H}^{-1}$ is bounded, and the spectral theorem can be used. However, in our problem $\hat{H}$ is not positive definite and hence $\hat{H}^{-1}$ does not have to be a bounded operator, and therefore in general the spectral theorem does not apply.

## D. The pseudo-ground

Let us consider a pair of oscillators associated with the complex frequencies $\omega_c$ and $\omega_c^*$. As seen previously, $\hat{H}_C$ is unbounded and has no eigenstates. However, one may still wish characterize the state wherein there is a minimal interaction between the moving bodies. In the Heisenberg picture, the quantized electromagnetic field is described by the field operator $\hat{\mathbf{F}}_C = \hat{\beta} e^{-i\omega_c t}\mathbf{f} + \hat{\chi} e^{-i\omega_c^* t}\mathbf{e} + \hat{\beta}^\dagger e^{i\omega_c^* t}\mathbf{f}^* + \hat{\chi}^\dagger e^{i\omega_c t}\mathbf{e}^*$ (for simplicity we consider only a pair of complex oscillators). In case of two weakly coupled moving bodies the theory of Sect. III applies, and hence it is possible to write $\mathbf{f}, \mathbf{e}$ in terms of the basis $\mathbf{F}_1, \mathbf{F}_2$ as in Eq. (22). Hence, at the time instant $t=0$ we may write:

$$\hat{\mathbf{F}}_C(t=0) = (\hat{\beta} + s\hat{\chi})\frac{\Omega_1}{|\Omega_1|}\frac{1}{\sqrt{2|E_{s,1}|}}\mathbf{F}_1 + (\hat{\beta} - s\hat{\chi})\frac{i}{\sqrt{2|E_{s,2}|}}\mathbf{F}_2 + h.c.. \qquad (59)$$

with $s = \text{sgn}(E_{s,1}) = -\text{sgn}(E_{s,2})$. Hence, the operators

$$\hat{E}_1 = \frac{1}{2}s(\hat{\beta} + s\hat{\chi})(\hat{\beta} + s\hat{\chi})^\dagger + h.c. \tag{60a}$$

$$\hat{E}_2 = -\frac{1}{2}s(\hat{\beta} - s\hat{\chi})(\hat{\beta} - s\hat{\chi})^\dagger + h.c. \tag{60b}$$

may be regarded to represent the wave energies stored in the first and second slab, respectively, at the instant $t = 0$. Assuming without loss of generality that $E_{s,1} > 0$, using Eq. (48), and taking into account that for weakly coupled slabs $\omega_c \approx \omega'$, we find that:

$$\hat{E}_1 = \frac{\hbar\omega'}{2}(\hat{a}_c\hat{a}_c^\dagger + \hat{a}_c^\dagger\hat{a}_c) \tag{61a}$$

$$\hat{E}_2 = -\frac{\hbar\omega'}{2}(\hat{b}_c\hat{b}_c^\dagger + \hat{b}_c^\dagger\hat{b}_c) \tag{61b}$$

The state that minimizes both $\hat{E}_1$ and $-\hat{E}_2$ is evidently the state where the occupation numbers vanish, $|0_c^a, 0_c^b\rangle$. Therefore, $|0_c^a, 0_c^b\rangle$ can be regarded as the state where the oscillations of the quantized fields are minimal at the time instant $t = 0$. It is important to emphasize that this property is only valid at $t = 0$ because we are considering the Heisenberg picture and the system properties vary with time. Thus, for instants $t \neq 0$ the state that corresponds to minimal oscillations of the fields is not $|0_c^a, 0_c^b\rangle$. It is relevant to mention that the operators $\hat{a}_c$ and $\hat{b}_c$ are understood as being always calculated at $t = 0$, because they are defined in terms of $\hat{\beta}, \hat{\chi}$ at $t = 0$ [Eq. (48)].

Even though the previous discussion assumed the scenario of two weakly coupled moving bodies, in the general case we may also regard $|0_c^a, 0_c^b\rangle$ as the state where the field oscillations at $t=0$ are minimal. Indeed, the strength of the field $\hat{\mathbf{F}}_C$ oscillations at $t=0$ may be described by the operator:

$$\begin{aligned}\hat{A}_C &= \frac{1}{2}\left(\hat{\beta}\hat{\beta}^\dagger + \hat{\beta}^\dagger\hat{\beta} + \hat{\chi}\hat{\chi}^\dagger + \hat{\chi}^\dagger\hat{\chi}\right) \\ &= \hat{\beta}\hat{\beta}^\dagger + \hat{\chi}\hat{\chi}^\dagger \\ &= \frac{\hbar|\omega_c|}{2}\left(\hat{a}_c^\dagger\hat{a}_c + \hat{b}_c^\dagger\hat{b}_c + 1\right)\end{aligned} \qquad (62)$$

Clearly, the quantum expectation of $\hat{A}_C$ is minimized for the state where the occupation numbers vanish, $\left|0_c^a, 0_c^b\right\rangle$.

The previous results can be extended in a trivial manner to the case of oscillators associated either with positive or negative real-valued frequencies. Thus, we introduce the state wherein all the occupation numbers (for oscillators associated with either real-valued or complex valued frequencies) vanish:

$$|\Omega\rangle = \left|\ldots,\left(0^a_{c,n\mathbf{k}}, 0^b_{c,n\mathbf{k}}\right),\ldots\right\rangle. \qquad (63)$$

This state should describe the situation where the field oscillations at $t=0$ are minimal, and is designated here by the pseudo-ground of the system at $t=0$. It is important to stress that $|\Omega\rangle$ is not a stationary state of the system and does not minimize the quantum expectation of the unbounded operator $\hat{H}_R + \hat{H}_C$. However, the analysis of Sect. III.B suggests that the total energy (including also the mechanical degrees of freedom) is minimal precisely when the wave fields vanish (because $dH_{tot,i}/dt > 0$ for the oscillations associated with growing exponentials, and $dH_{tot,i}/dt < 0$ for the oscillations associated with decaying exponentials). Thus the pseudo-ground may be regarded as the state that minimizes the total energy ($H_{tot}$) when the system is subject to the constraint $v_i = const_i..$

## VI. Time evolution of the pseudo-ground state

In order to apply the developed quantum theory, we study the time evolution of the pseudo-ground state $|\Omega\rangle$. For simplicity, we consider a single pair of oscillators associated with the complex frequencies $\omega_c = \omega' + i\lambda$ and $\omega_c^*$, such that the Hamiltonian of the system is (compare with Eq. (50)):

$$\hat{H} = \frac{\hbar \omega'}{2}\left(\hat{a}_c \hat{a}_c^\dagger + \hat{a}_c^\dagger \hat{a}_c - \hat{b}_c \hat{b}_c^\dagger - \hat{b}_c^\dagger \hat{b}_c\right) + i\hbar\lambda\left(\hat{a}_c^\dagger \hat{b}_c^\dagger - \hat{a}_c \hat{b}_c\right). \tag{64}$$

Our objective it to study the time evolution of the state $\left|0_c^a, 0_c^b\right\rangle$. It is interesting to note that the above Hamiltonian can be regarded as being due to the interaction of a quantum harmonic oscillator with positive frequency ($\frac{\hbar\omega'}{2}\left(\hat{a}_c \hat{a}_c^\dagger + \hat{a}_c^\dagger \hat{a}_c\right)$) and a quantum harmonic oscillator with negative frequency $-\frac{\hbar\omega'}{2}\left(\hat{b}_c \hat{b}_c^\dagger + \hat{b}_c^\dagger \hat{b}_c\right)$. The interaction term is $\hat{H}_{int} = i\hbar\lambda\left(\hat{a}_c^\dagger \hat{b}_c^\dagger - \hat{a}_c \hat{b}_c\right)$. It is also relevant to mention that this interaction term can be regarded as the rotating wave approximation of a more general interaction of the type $i\hbar\lambda\left(\hat{a}_c^\dagger + \hat{a}_c\right)\left(\hat{b}_c^\dagger - \hat{b}_c\right)$ (note that $\hat{a}_c^\dagger$ creates a particle with positive energy and $\hat{b}_c^\dagger$ creates a particle with negative energy).

We look for a solution of the Schrödinger equation of the form $|\psi\rangle = \sum_{n=0}^{\infty} c_n |n,n\rangle$. It is simple to check that $\hat{H}|\psi\rangle = i\hbar\lambda \sum_{n=0}^{\infty} [c_n(n+1)|n+1,n+1\rangle - c_n n|n-1,n-1\rangle]$ and thus the coefficients $c_n$ are required to satisfy:

$$\frac{dc_n}{dt} = \lambda[nc_{n-1} - (n+1)c_{n+1}], \qquad n \geq 0. \tag{65}$$

In order that the initial state is the pseudo-ground $|0_c^a, 0_c^b\rangle$, we must also impose that $c_n(t=0) = \delta_{n,0}$. This infinite system of ordinary differential equations has an analytical solution. It can be checked by direct substitution that the solution is exactly:

$$c_n(t) = \text{sech}(t\lambda)[\tanh(t\lambda)]^n, \qquad n \geq 0. \tag{66}$$

As it should be, the normalization condition $\sum_{n=0}^{\infty} |c_n(t)|^2 = 1$ holds at all time instants. In Fig. 4 we depict some of the coefficients $c_n(t)$ as a function of the normalized time.

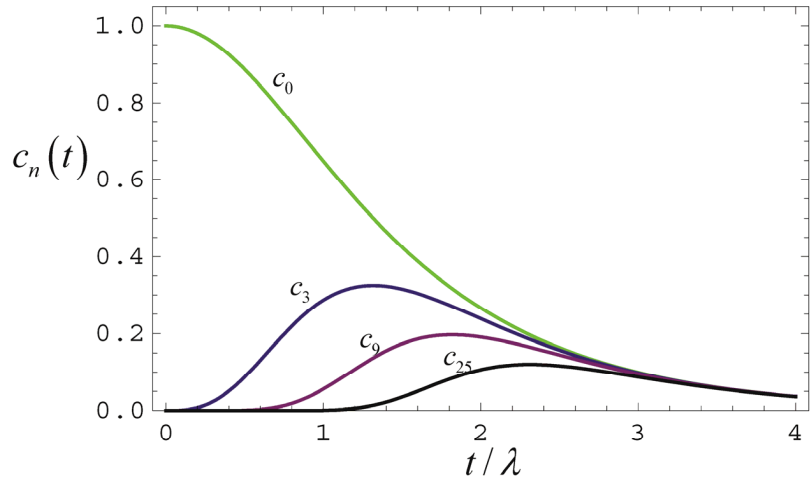

Fig. 4. (Color online) Coefficients $c_n(t)$ (for n=0, 3, 9, 25) as a function of the normalized time, assuming that the system is prepared in the initial state $\left|0_c^a, 0_c^b\right\rangle$, i.e. $c_n(t=0) = \delta_{n,0}$.

As seen, as the time passes $c_0(t) \to 0$, and the probability of occupation of the states $|n,n\rangle$ becomes nonzero. It is interesting to note that for sufficiently large $t$ the coefficients $c_n(t)$ approach $c_n(t) \approx \mathrm{sech}(t\lambda)$. Thus, for a given $n_0$ and for sufficiently large $t$ the states $|n,n\rangle$ with $n \leq n_0$ have nearly the same probability of being occupied. This property is also seen in the plots of Fig. 4. This result shows that a system prepared in the pseudo-ground, i.e. in the state wherein the fluctuations of the fields are minimal, evolves with time in such a manner that the fluctuations of the fields become larger and larger and the probability of finding the system in states with large occupation numbers (let us say in states with $n > n_0$) is larger and larger. However, the total wave energy of the system is preserved and is identical to zero. These properties are consistent with the picture provided by a classical framework (see Fig. 2).

## VII. Conclusion

In this work, we investigated the natural modes of oscillation of the electromagnetic field in a system of moving bodies with time independent velocities. It was shown that above the Cherenkov threshold the wave energy may become negative, but the total system energy is always positive. We studied the hybridization of the guided modes supported by two material slabs in the limit of a weak coupling, showing that when the wave energies of the guided modes have different signs the hybridized modes may be characterized by a complex valued frequency of oscillation, leading to oscillations of the fields with infinitely large amplitude in the absence of nonlinear effects. We developed a theory to quantize the electromagnetic field in a generic system of moving bodies allowing for velocities exceeding the Cherenkov threshold. The Hamiltonian of the system is written in terms of quantum harmonic oscillators associated with either positive or negative frequencies. The oscillators associated with complex valued oscillation frequencies are coupled by an interaction term that does not preserve the number of quanta. The quantized field operators satisfy the Maxwell's equations and standard canonical commutation relations. In general, the quantized Hamiltonian does not have stationary states and has no ground state. However, it is possible to introduce a pseudo-ground state where the oscillation amplitudes are minimal. It was shown that a system prepared in the pseudo-ground state evolves in such a manner that states with large occupation numbers have the same probability of being occupied as states with small occupation numbers. The developed theory assumes that the velocity of the moving slabs is independent of time. However, the exchange of wave energy between the different bodies is accompanied by an exchange of wave momentum, and thus an external action is required to keep the kinetic momentum of the slabs independent of time. These ideas are further explored elsewhere [20].

**Acknowledgements:** This work is supported in part by Fundação para a Ciência e a Tecnologia grant number PTDC/EEI-TEL/2764/2012.

## Appendix A: The force associated with the stress tensor

In what follows, Eq. (12) of the main text is demonstrated. We start by noting that the following identity holds for arbitrary vectors:

$$\hat{\mathbf{x}} \cdot (\mathbf{E} \nabla \cdot \mathbf{D} - \mathbf{D} \times \nabla \times \mathbf{E}) = \nabla \cdot (\mathbf{D} E_x) - \mathbf{D} \cdot \frac{\partial \mathbf{E}}{\partial x}. \tag{A1}$$

Thus, supposing that **D** and **E** fields satisfy the source-free macroscopic Maxwell equations, we obtain after some manipulations that

$$\begin{aligned}\hat{\mathbf{x}} \cdot \left( \mathbf{D} \times \frac{\partial \mathbf{B}}{\partial t} \right) &= \nabla \cdot (\mathbf{D} E_x) - \mathbf{D} \cdot \frac{\partial \mathbf{E}}{\partial x} \\ &= \nabla \cdot \left( \mathbf{D} E_x - \tfrac{1}{2} \varepsilon_0 \mathbf{E} \cdot \mathbf{E} \hat{\mathbf{x}} \right) - \mathbf{P}_e \cdot \frac{\partial \mathbf{E}}{\partial x}\end{aligned} \tag{A2}$$

where $\mathbf{P}_e = \mathbf{D} - \varepsilon_0 \mathbf{E}$ is the electric polarization vector. Proceeding in the same manner for the magnetic field, using $\hat{\mathbf{x}} \cdot (\mathbf{H} \nabla \cdot \mathbf{B} - \mathbf{B} \times \nabla \times \mathbf{H}) = \nabla \cdot (\mathbf{B} H_x) - \mathbf{B} \cdot \frac{\partial \mathbf{H}}{\partial x}$, it follows that

$$\hat{\mathbf{x}} \cdot \left( \frac{\partial \mathbf{D}}{\partial t} \times \mathbf{B} \right) = \nabla \cdot \left( \mathbf{B} H_x - \tfrac{1}{2} \mu_0 \mathbf{H} \cdot \mathbf{H} \hat{\mathbf{x}} \right) - \mathbf{P}_m \cdot \frac{\partial \mathbf{H}}{\partial x}, \tag{A3}$$

where $\mathbf{P}_m = \mathbf{B} - \mu_0 \mathbf{H}$ is the magnetic polarization vector. Combining Eqs. (A2) and (A3) it is found that:

$$\mathbf{P}_e \cdot \frac{\partial \mathbf{E}}{\partial x} + \mathbf{P}_m \cdot \frac{\partial \mathbf{H}}{\partial x} = \nabla \cdot \left( \mathbf{D} E_x - \tfrac{1}{2} \varepsilon_0 \mathbf{E} \cdot \mathbf{E} \hat{\mathbf{x}} + \mathbf{B} H_x - \tfrac{1}{2} \mu_0 \mathbf{H} \cdot \mathbf{H} \hat{\mathbf{x}} \right) - \frac{\partial}{\partial t} (\mathbf{D} \times \mathbf{B}) \cdot \hat{\mathbf{x}} \tag{A4}$$

Next, we note that the left-hand side of the above equation can be written as:

$$\mathbf{P}_e \cdot \frac{\partial \mathbf{E}}{\partial x} + \mathbf{P}_m \cdot \frac{\partial \mathbf{H}}{\partial x} = \frac{\partial \mathbf{F}}{\partial x} \cdot (\mathbf{G} - \mathbf{M}_0 \cdot \mathbf{F})$$
$$= \frac{\partial \mathbf{F}}{\partial x} \cdot (\mathbf{M} - \mathbf{M}_0) \cdot \mathbf{F} \tag{A5}$$

where $\mathbf{F}$ and $\mathbf{G}$ are the six-vectors defined in the main text, $\mathbf{M}$ is the material matrix and $\mathbf{M}_0$ is the material matrix for a vacuum (a diagonal matrix with entries $\varepsilon_0, \mu_0$ in the main diagonal). We are interested in systems that are invariant to translations along the $x$-direction, such that $\mathbf{M}$ is independent of $x$. In these cases, taking into account that $\mathbf{M} - \mathbf{M}_0$ is a symmetric matrix we can write

$$\mathbf{P}_e \cdot \frac{\partial \mathbf{E}}{\partial x} + \mathbf{P}_m \cdot \frac{\partial \mathbf{H}}{\partial x} = \frac{1}{2} \frac{\partial}{\partial x} \left[ \mathbf{F} \cdot (\mathbf{M} - \mathbf{M}_0) \cdot \mathbf{F} \right]. \tag{A6}$$

Next, the above expression is substituted into Eq. (A4) and both sides of the resulting equation are integrated over the volume $V_i$ of a given body, including the boundary surface. Taking into account that the walls of the cavity that encloses the body are terminated with periodic boundary conditions, it is seen that $\int_{V_i} \frac{1}{2} \frac{\partial}{\partial x} \left[ \mathbf{F} \cdot (\mathbf{M} - \mathbf{M}_0) \cdot \mathbf{F} \right] d^3\mathbf{r} = 0$. Thus, it follows that:

$$\frac{dp_{w,i}}{dt} = \int_{V_i} \nabla \cdot \left( \mathbf{D} E_x - \frac{1}{2} \varepsilon_0 \mathbf{E} \cdot \mathbf{E} \hat{\mathbf{x}} + \mathbf{B} H_x - \frac{1}{2} \mu_0 \mathbf{H} \cdot \mathbf{H} \hat{\mathbf{x}} \right) d^3\mathbf{r} \tag{A7}$$

where we introduced the wave momentum $p_{w,i} = \int_{V_i} (\mathbf{D} \times \mathbf{B}) \cdot \hat{\mathbf{x}} \, d^3\mathbf{r}$ of the $i$-th body. Using the divergence theorem, the volume integral can be written in terms of the Maxwell stress tensor in the vacuum region,

$$\frac{dp_{w,i}}{dt} = \int_{\partial V_i} \hat{\mathbf{n}} \cdot \overline{\overline{\mathbf{T}}} \cdot \hat{\mathbf{x}} \, ds, \tag{A8}$$

where $\partial V_i$ the boundary of $V_i$ (contained in the vacuum region), and

$$\overline{\overline{\mathbf{T}}} = \varepsilon_0 \mathbf{E} \otimes \mathbf{E} - \frac{1}{2}\varepsilon_0 \mathbf{E} \cdot \mathbf{E} \overline{\overline{\mathbf{I}}} + \frac{1}{\mu_0} \mathbf{B} \otimes \mathbf{B} - \frac{1}{2\mu_0} \mathbf{B} \cdot \mathbf{B} \overline{\overline{\mathbf{I}}}$$ is the stress tensor in the vacuum region.

Finally, substituting Eq. (A8) into Eq. (11) we obtain the desired result.

## Appendix B: Proof that $\Omega_1 / \Omega_2^* = E_{s,1} / E_{s,2}$

Here, we prove that for the system analyzed in Sect. III the coupling constants are such that $\Omega_1 / \Omega_2^* = E_{s,1} / E_{s,2}$. To this end, we note that from Eq. (15) it follows that:

$$\begin{aligned}\left\langle \mathbf{F}_2 \mid \hat{N} \mid \mathbf{F}_1 \right\rangle_c &= \omega' \left\langle \mathbf{F}_2 \mid \mathbf{M}_1 \mid \mathbf{F}_1 \right\rangle_c \\ &= \omega' \left\langle \mathbf{F}_2 \mid \mathbf{M}_0 \mid \mathbf{F}_1 \right\rangle_c + \left\langle \mathbf{F}_2 \mid \mathbf{j}_1 \right\rangle_c \end{aligned} \quad (B1)$$

where $\mathbf{j}_1 = \omega' (\mathbf{M}_1 - \mathbf{M}_0) \cdot \mathbf{F}_1$ and $\mathbf{M}_0$ represents the material matrix for the vacuum. An analogous formula holds if the subscripts "1" and "2" are interchanged. Hence, it is possible to write $\left\langle \mathbf{F}_2 \mid \hat{N} - \omega' \mathbf{M}_0 \mid \mathbf{F}_1 \right\rangle_c = \left\langle \mathbf{F}_2 \mid \mathbf{j}_1 \right\rangle_c$ and $\left\langle \mathbf{F}_1 \mid \hat{N} - \omega' \mathbf{M}_0 \mid \mathbf{F}_2 \right\rangle_c = \left\langle \mathbf{F}_1 \mid \mathbf{j}_2 \right\rangle_c$. Because the operator $\hat{N} - \omega' \mathbf{M}_0$ is Hermitian with respect to the canonical inner product, it follows that:

$$\left\langle \mathbf{F}_2 \mid \mathbf{j}_1 \right\rangle_c = \left\langle \mathbf{F}_1 \mid \mathbf{j}_2 \right\rangle_c^*. \quad (B2)$$

Next, we note that because $\mathbf{M}_1 - \mathbf{M}_0 = \mathbf{M} - \mathbf{M}_2$ it follows that:

$$\left\langle \mathbf{F}_2 \mid \mathbf{M} - \mathbf{M}_2 \mid \mathbf{F}_1 \right\rangle_c = \left\langle \mathbf{F}_1 \mid \mathbf{M} - \mathbf{M}_1 \mid \mathbf{F}_2 \right\rangle_c^*. \quad (B3)$$

Thus, it is evident from the definition of $\Omega_1$ and $\Omega_2$ that they satisfy $\Omega_1 / \Omega_2^* = E_{s,1} / E_{s,2}$, as we wanted to show.

## Appendix C: The sign of $v_{ph,i} v_i > 0$

Let $(\mathbf{E}_l, \mathbf{B}_l)$ and $(\mathbf{D}_l, \mathbf{H}_l)$ with ($l$=1,2) be arbitrary solutions of the source-free macroscopic Maxwell's equations in the lab frame. Define $(\mathbf{E}'_l, \mathbf{B}'_l)$ and $(\mathbf{D}'_l, \mathbf{H}'_l)$ as the corresponding relativistically transformed fields calculated in the reference frame that moves with velocity $\mathbf{v} = v\hat{\mathbf{x}}$ with respect to the lab frame (see Ref. [22, Sect. 11.10]; the $(c\mathbf{D}, \mathbf{H})$ fields are transformed in the same manner as the $(\mathbf{E}, c\mathbf{B})$ fields). Explicit calculations show that if we put:

$$L_d = \frac{1}{4}\left(\mathbf{D}_1 \cdot \mathbf{E}_2 + \mathbf{D}_2 \cdot \mathbf{E}_1 - \mathbf{B}_1 \cdot \mathbf{H}_2 - \mathbf{B}_2 \cdot \mathbf{H}_1\right) \tag{C1}$$

$$g_{ps,x} = \frac{1}{2}\hat{\mathbf{x}} \cdot \left[\mathbf{D}_1 \times \mathbf{B}_2 + \mathbf{D}_2 \times \mathbf{B}_1 - \frac{1}{c^2}\left(\mathbf{E}_1 \times \mathbf{H}_2 + \mathbf{E}_2 \times \mathbf{H}_1\right)\right], \tag{C2}$$

then

$$g_{ps,x} = g'_{ps,x}, \quad \text{and} \quad L_d = L'_d \tag{C3}$$

where $g'_{ps,x}$ and $L'_d$ are calculated with the primed fields, and the two sides of the previous equations are evaluated at corresponding spacetime points. Suppose now that we choose $\mathbf{F}_2$ as the field associated with index $l$=1 and $\mathbf{F}_2^*$ as the field associated with index $l$=2, being $\mathbf{F}_2$ defined as in Sect. III. Moreover suppose that the primed frame is co-moving with the body 2. Then, from Eqs. (25), (26), and $g_{ps,x} = g'_{ps,x}$ it follows that:

$$\frac{1}{v_{ph,2}}\left(1 - \frac{v_{ph,2}v_{g,2}}{c^2}\right)\frac{E_{s,2}}{V_2} = \frac{1}{v_{ph,2}^{co}}\left(1 - \frac{v_{ph,2}^{co}v_{g,2}^{co}}{c^2}\right)\left(\frac{E_{s,2}}{V_2}\right)^{co} \tag{C4}$$

where $V_2$ represents the volume of the body. The superscript "co" indicates that the pertinent quantity is calculated in the co-moving frame. Let us suppose next that $E_{s,2} < 0$ in the lab frame and that $k_y = 0$. Because $E_{s,2}^{co}$ is necessarily positive, and because for a waveguide made of a nondispersive dielectric one has $\left|v_{ph,i}^{co}\right| < c/n$ and $\left|v_{g,i}^{co}\right| < c/n$ when $k_y = 0$, it follows that:

$$v_{ph,2} v_{ph,2}^{co} < 0, \qquad \text{if } E_{s,2} < 0 \text{ and } k_y = 0. \tag{C5}$$

Hence, taking into account that $v_{ph,i}^{co} = \dfrac{v_{ph,i} - v_i}{1 - v_{ph,i} v_i / c^2}$ it can be easily checked that Eq. (C5) can hold only if $v_{ph,2} v_2 > 0$. Therefore, we have shown that if $E_{s,2} < 0$ and $k_y = 0$ then $v_{ph,2} v_2 > 0$.

## Appendix D: The proper sub-spaces of the operator $\mathbf{M}^{-1} \cdot \hat{N}$

In this Appendix, we derive some key properties of the proper-subspaces of the operator $\mathbf{M}^{-1} \cdot \hat{N}$. We start by noting that the inner product $\langle \ | \ \rangle$ is non-degenerate, i.e. there is no $\mathbf{F} \neq 0$ orthogonal to every element of the space, or equivalently such that the 1-form $\langle \mathbf{F} | \ . \rangle$ is identically zero. The reason is simple: the material matrix $\mathbf{M}$ is assumed to have $\det(\mathbf{M}) \neq 0$ in every point of space.

As a consequence of this property, under the hypothesis that that every element in the space can be written as a linear combination of eigenvectors $\mathbf{M}^{-1} \cdot \hat{N}$, it follows that the restriction of $\langle \ | \ \rangle$ to the sub-spaces $S_{\omega_i}$, and $S_{\omega_{c,i}, \omega_{c,i}^*}$ is also non-degenerate. Indeed, if there was an $\mathbf{F} \neq 0$ in $S_{\omega_i}$ orthogonal to every element in $S_{\omega_i}$, then, taking into account that Eq. (30) establishes a decomposition of the space into orthogonal subspaces, it would follow that $\mathbf{F}$ would be orthogonal to every element of the space, which as discussed previously is not possible. The same argument can be used to show that the restriction of $\langle \ | \ \rangle$ to $S_{\omega_{c,i},\omega_{c,i}^*}$ is non-degenerate.

The quadratic form associated with the inner product $\langle \ | \ \rangle$ is represented in the proper sub-space $S_{\omega_i}$ by a Hermitian matrix. This matrix cannot be singular because $\langle \ | \ \rangle$ is non-degenerate. This result implies that it is possible to find an orthogonal basis $\{\mathbf{F}_n\}$ of $S_{\omega_i}$ such:

$$\langle \mathbf{F}_n | \mathbf{F}_m \rangle = \pm \delta_{m,n}, \qquad \text{(basis of } S_{\omega_i}) \tag{D1}$$

Note that because the inner product $\langle \ | \ \rangle$ is indefinite the normalized wave energy $\langle \mathbf{F}_n | \mathbf{F}_n \rangle$ can be either positive or negative.

Next, we characterize the eigenspace $S_{\omega_{c,i},\omega_{c,i}^*}$. To this end we start by considering a basis with elements $\mathbf{f}_n, \mathbf{e}_n$, $n=1,2,\ldots,N$ with $\mathbf{f}_n$ an eigenfunction associated with the complex-valued eigenfrequencies $\omega_{c,n}$ with $\lambda_n = \text{Im}\{\omega_n\} > 0$, and with $\mathbf{e}_n = \tilde{\mathbf{f}}_n$ being dual of $\mathbf{f}_n$ [defined as in Eq. (34)], which is associated with the eigenfrequency $\omega_{c,n}^*$. Writing $\mathbf{F} = \sum \beta_n \mathbf{f}_n + \chi_n \mathbf{e}_n$, one sees that the quadratic form associated with the inner product $\langle \ | \ \rangle$ is represented by

$$\langle \mathbf{F} | \mathbf{F} \rangle = \begin{pmatrix} \beta & \chi \end{pmatrix}^* \cdot \begin{pmatrix} 0 & \mathbf{A} \\ \mathbf{B} & 0 \end{pmatrix} \cdot \begin{pmatrix} \beta \\ \chi \end{pmatrix}, \tag{D2}$$

with $\mathbf{A} = \left[\langle \mathbf{f}_m | \mathbf{e}_n \rangle\right]_{N \times N}$ and $\mathbf{B} = \left[\langle \mathbf{e}_m | \mathbf{f}_n \rangle\right]_{N \times N} = \mathbf{A}^\dagger$ square matrices. Because of the property (35c) the matrix $\mathbf{A}$ is symmetric and therefore $\mathbf{B} = \mathbf{A}^*$. Moreover, because the restriction of $\langle \, | \, \rangle$ to $S_{\omega_{c,i},\omega_{c,i}^*}$ is non-degenerate, $\mathbf{A}$ must be invertible. Tagaki's decomposition states that a complex-valued symmetric invertible matrix can be decomposed as $\mathbf{V} \cdot \mathbf{D} \cdot \mathbf{V}^t$, where $\mathbf{D}$ is a diagonal non-negative matrix and $\mathbf{V}$ is unitary. This result implies that $\mathbf{A}$ can be factorized as $\mathbf{A} = \mathbf{S}^t \cdot \mathbf{S}$ for some invertible matrix $\mathbf{S}$. Therefore, we may rewrite Eq. (D2) as,

$$\langle \mathbf{F} | \mathbf{F} \rangle = \begin{pmatrix} \beta' & \chi' \end{pmatrix}^* \cdot \begin{pmatrix} 0 & \mathbf{I} \\ \mathbf{I} & 0 \end{pmatrix} \cdot \begin{pmatrix} \beta' \\ \chi' \end{pmatrix}, \qquad \text{with} \quad \begin{pmatrix} \beta' \\ \chi' \end{pmatrix} = \begin{pmatrix} \mathbf{S}^* & 0 \\ 0 & \mathbf{S} \end{pmatrix} \begin{pmatrix} \beta \\ \chi \end{pmatrix} \qquad (\text{D3})$$

where $\mathbf{I}$ represents the $N \times N$ identity matrix. Hence, it follows that it is possible to find a basis with elements $\mathbf{f}_n', \mathbf{e}_n'$, $n=1,2,\ldots,N$ such that:

$$\langle \mathbf{e}_m' | \mathbf{f}_n' \rangle = \delta_{m,n} \qquad ; \qquad \langle \mathbf{f}_m' | \mathbf{f}_n' \rangle = \langle \mathbf{e}_m' | \mathbf{e}_n' \rangle = 0 \qquad (\text{D4})$$

Using Eq. (D3) and (35b) it is simple to check that $\mathbf{e}_m' = \tilde{\mathbf{f}}_m'$.

## Appendix E: The commutation relations for the field operators

Next, we derive the equal time commutation relations for the quantized fields. We start by noting that from Eqs. (52)-(53), and from the commutation relations (47) and (51), it is possible to write:

$$\begin{aligned}
\left[\hat{\mathbf{G}}(\mathbf{r}), \hat{\mathbf{G}}(\mathbf{r}')\right] &= \left[\mathbf{M} \cdot \hat{\mathbf{F}}_R(\mathbf{r}), \mathbf{M} \cdot \hat{\mathbf{F}}_R(\mathbf{r}')\right] + \left[\mathbf{M} \cdot \hat{\mathbf{F}}_C(\mathbf{r}), \mathbf{M} \cdot \hat{\mathbf{F}}_C(\mathbf{r}')\right] \\
&= \sum_{n\mathbf{k} \in E_R} \frac{\hbar |\omega_{n\mathbf{k}}|}{2} \left\{ \mathbf{G}_{n\mathbf{k}}(\mathbf{r}) \otimes \mathbf{G}_{n\mathbf{k}}^*(\mathbf{r}') - \mathbf{G}_{n\mathbf{k}}^*(\mathbf{r}) \otimes \mathbf{G}_{n\mathbf{k}}(\mathbf{r}') \right\} \\
&\quad + \sum_{n\mathbf{k} \in E_C} \frac{\hbar \omega_{c,n\mathbf{k}}}{2} \left\{ \mathbf{g}_{n\mathbf{k}}(\mathbf{r}) \otimes \tilde{\mathbf{g}}_{n\mathbf{k}}^*(\mathbf{r}') - \tilde{\mathbf{g}}_{n\mathbf{k}}^*(\mathbf{r}) \otimes \mathbf{g}_{n\mathbf{k}}(\mathbf{r}') \right\} \\
&\quad + \frac{\hbar \omega_{c,n\mathbf{k}}^*}{2} \left\{ \tilde{\mathbf{g}}_{n\mathbf{k}}(\mathbf{r}) \otimes \mathbf{g}_{n\mathbf{k}}^*(\mathbf{r}') - \mathbf{g}_{n\mathbf{k}}^*(\mathbf{r}) \otimes \tilde{\mathbf{g}}_{n\mathbf{k}}(\mathbf{r}') \right\}
\end{aligned} \qquad (\text{E1})$$

where $\mathbf{G}_{n\mathbf{k}} = \mathbf{M} \cdot \mathbf{F}_{n\mathbf{k}}$, $\mathbf{g}_{n\mathbf{k}} = \mathbf{M} \cdot \mathbf{f}_{n\mathbf{k}}$, $\tilde{\mathbf{g}}_{n\mathbf{k}} = \mathbf{M} \cdot \mathbf{e}_{n\mathbf{k}}$, and $\otimes$ represents the tensor product of two vectors.

Supposing that the eigenfunctions of $\mathbf{M}^{-1} \cdot \hat{N}$ span the space of six-vectors with periodic boundary conditions, it follows that it is possible to expand any vector function in terms of the basis formed by $\mathbf{F}_{n\mathbf{k}}$, $\mathbf{f}_{n\mathbf{k}}$, $\mathbf{e}_{n\mathbf{k}}$, and $\mathbf{L}_{n\mathbf{k}}$, being $\mathbf{L}_{n\mathbf{k}}$ a basis of the null-space of $\mathbf{M}^{-1} \cdot \hat{N}$, i.e. $\left(\mathbf{M}^{-1} \cdot \hat{N}\right)\mathbf{L}_{n\mathbf{k}} = 0$ (note that $S_{\mathbf{M}^{-1} \cdot \hat{N}}$ defined by Eq. (30) is the subspace generated by the eigenfunctions associated with nonzero eigenvalues). In particular, the real valued function $\mathbf{V}\delta(\mathbf{r} - \mathbf{r}')$ ($\mathbf{V}$ is a generic real valued six-vector), can be expanded as follows:

$$\mathbf{V}\delta(\mathbf{r} - \mathbf{r}') = \sum_{n\mathbf{k} \in E_R \cup E_0} \left(\alpha_{n\mathbf{k}} \mathbf{F}_{n\mathbf{k}} + \alpha_{n\mathbf{k}}^* \mathbf{F}_{n\mathbf{k}}^*\right) + \sum_{n\mathbf{k} \in E_C} \left(\beta_{n\mathbf{k}} \mathbf{f}_{n\mathbf{k}} + \chi_{n\mathbf{k}} \mathbf{e}_{n\mathbf{k}} + \beta_{n\mathbf{k}}^* \mathbf{f}_{n\mathbf{k}}^* + \chi_{n\mathbf{k}}^* \mathbf{e}_{n\mathbf{k}}^*\right) \quad (E2)$$

The above expansion is formally equivalent to Eq. (39), except that summation range in the first sum is extended to the set $E_0 = \{(n, \mathbf{k}): \omega_{n\mathbf{k}} = 0 \text{ and } k_x > 0\}$. The eigenfunctions $\mathbf{F}_{n\mathbf{k}}$ associated with the set $E_0$ should be identified with the eigenfunctions $\mathbf{L}_{n\mathbf{k}}$. Using the orthogonality conditions (38) (which apply also to the extended set of eigenfunctions), one easily finds that

$$\alpha_{n\mathbf{k}} \langle \mathbf{F}_{n\mathbf{k}} | \mathbf{F}_{n\mathbf{k}} \rangle = \langle \mathbf{F}_{n\mathbf{k}} | \mathbf{V}\delta(\mathbf{r} - \mathbf{r}') \rangle = \frac{1}{2} \mathbf{G}_{n\mathbf{k}}^*(\mathbf{r}') \cdot \mathbf{V} \quad (E3a)$$

$$\beta_{n\mathbf{k}} = \langle \mathbf{e}_{n\mathbf{k}} | \mathbf{V}\delta(\mathbf{r} - \mathbf{r}') \rangle = \frac{1}{2} \tilde{\mathbf{g}}_{n\mathbf{k}}^*(\mathbf{r}') \cdot \mathbf{V} \quad (E3b)$$

$$\chi_{n\mathbf{k}} = \langle \mathbf{f}_{n\mathbf{k}} | \mathbf{V}\delta(\mathbf{r} - \mathbf{r}') \rangle = \frac{1}{2} \mathbf{g}_{n\mathbf{k}}^*(\mathbf{r}') \cdot \mathbf{V} \quad (E3c)$$

Hence, substituting these formulas into Eq. (E2), and taking into account that $\mathbf{V}$ is a generic real valued six-vector, it follows that

$$\mathbf{I}_{6\times 6}\delta(\mathbf{r}-\mathbf{r}') = \sum_{n\mathbf{k}\in E_R\cup E_0}\frac{1}{2\langle \mathbf{F}_{n\mathbf{k}}|\mathbf{F}_{n\mathbf{k}}\rangle}\left(\mathbf{F}_{n\mathbf{k}}(\mathbf{r})\otimes\mathbf{G}^*_{n\mathbf{k}}(\mathbf{r}')+\mathbf{F}^*_{n\mathbf{k}}(\mathbf{r})\otimes\mathbf{G}_{n\mathbf{k}}(\mathbf{r}')\right)$$
$$+\frac{1}{2}\sum_{n\mathbf{k}\in E_C}\{\mathbf{f}_{n\mathbf{k}}(\mathbf{r})\otimes\tilde{\mathbf{g}}^*_{n\mathbf{k}}(\mathbf{r}')+\mathbf{f}^*_{n\mathbf{k}}(\mathbf{r})\otimes\tilde{\mathbf{g}}_{n\mathbf{k}}(\mathbf{r}')+ \quad\quad (E4)$$
$$+\mathbf{e}_{n\mathbf{k}}(\mathbf{r})\otimes\mathbf{g}^*_{n\mathbf{k}}(\mathbf{r}')+\mathbf{e}^*_{n\mathbf{k}}(\mathbf{r})\otimes\mathbf{g}_{n\mathbf{k}}(\mathbf{r}')\}$$

where $\mathbf{I}_{6\times 6}$ represents the identity tensor and $\langle \mathbf{F}_{n\mathbf{k}}|\mathbf{F}_{n\mathbf{k}}\rangle = \pm 1$. Applying the operator $\hat{N}$ to the left-hand side of both members, and using Eqs. (4) and (32), it follows that:

$$\hat{N}\cdot\{\mathbf{I}_{6\times 6}\delta(\mathbf{r}-\mathbf{r}')\} = \sum_{n\mathbf{k}\in E_R\cup E_0}\frac{\omega_{n\mathbf{k}}}{2\langle \mathbf{F}_{n\mathbf{k}}|\mathbf{F}_{n\mathbf{k}}\rangle}\left(\mathbf{G}_{n\mathbf{k}}(\mathbf{r})\otimes\mathbf{G}^*_{n\mathbf{k}}(\mathbf{r}')-\mathbf{G}^*_{n\mathbf{k}}(\mathbf{r})\otimes\mathbf{G}_{n\mathbf{k}}(\mathbf{r}')\right)$$
$$+\frac{1}{2}\sum_{n\mathbf{k}\in E_C}\{\omega_{c,n\mathbf{k}}\mathbf{g}_{n\mathbf{k}}(\mathbf{r})\otimes\tilde{\mathbf{g}}^*_{n\mathbf{k}}(\mathbf{r}')-\omega^*_{c,n\mathbf{k}}\mathbf{g}^*_{n\mathbf{k}}(\mathbf{r})\otimes\tilde{\mathbf{g}}_{n\mathbf{k}}(\mathbf{r}')+ \quad\quad (E5)$$
$$+\omega^*_{c,n\mathbf{k}}\tilde{\mathbf{g}}_{n\mathbf{k}}(\mathbf{r})\otimes\mathbf{g}^*_{n\mathbf{k}}(\mathbf{r}')-\omega_{c,n\mathbf{k}}\tilde{\mathbf{g}}^*_{n\mathbf{k}}(\mathbf{r})\otimes\mathbf{g}_{n\mathbf{k}}(\mathbf{r}')\}$$

Noting that for indices in the set $E_0$ the eigenfrequencies $\omega_{n\mathbf{k}}$ vanish, whereas for indices in the set $E_R$ one has $\frac{\omega_{n\mathbf{k}}}{\langle \mathbf{F}_{n\mathbf{k}}|\mathbf{F}_{n\mathbf{k}}\rangle} = |\omega_{n\mathbf{k}}|$, it is found that:

$$\hat{N}\cdot\{\mathbf{I}_{6\times 6}\delta(\mathbf{r}-\mathbf{r}')\} = \sum_{n\mathbf{k}\in E_R}\frac{|\omega_{n\mathbf{k}}|}{2}\left(\mathbf{G}_{n\mathbf{k}}(\mathbf{r})\otimes\mathbf{G}^*_{n\mathbf{k}}(\mathbf{r}')-\mathbf{G}^*_{n\mathbf{k}}(\mathbf{r})\otimes\mathbf{G}_{n\mathbf{k}}(\mathbf{r}')\right)$$
$$+\sum_{n\mathbf{k}\in E_C}\frac{\omega_{c,n\mathbf{k}}}{2}\{\mathbf{g}_{n\mathbf{k}}(\mathbf{r})\otimes\tilde{\mathbf{g}}^*_{n\mathbf{k}}(\mathbf{r}')-\tilde{\mathbf{g}}^*_{n\mathbf{k}}(\mathbf{r})\otimes\mathbf{g}_{n\mathbf{k}}(\mathbf{r}')\} \quad\quad (E6)$$
$$+\frac{\omega^*_{c,n\mathbf{k}}}{2}\{\tilde{\mathbf{g}}_{n\mathbf{k}}(\mathbf{r})\otimes\mathbf{g}^*_{n\mathbf{k}}(\mathbf{r}')-\mathbf{g}^*_{n\mathbf{k}}(\mathbf{r})\otimes\tilde{\mathbf{g}}_{n\mathbf{k}}(\mathbf{r}')\}$$

Comparing this result with Eq. (E1), we finally find that $\left[\hat{\mathbf{G}}(\mathbf{r}),\hat{\mathbf{G}}(\mathbf{r}')\right] = \hbar\hat{N}\cdot\{\mathbf{I}_{6\times 6}\delta(\mathbf{r}-\mathbf{r}')\}$.